%% file: paper.tex
\pdfoutput=1

\newif\ifconfversion
\confversionfalse

\documentclass[letterpaper,twocolumn,10pt]{article}
\PassOptionsToPackage{hyphens}{url}
\usepackage[style=base]{caption}
\ifconfversion
\usepackage[10pt]{sigmin}
\else
\usepackage[10pt,nocopyright]{sigmin}
\fi
\usepackage[breaklinks, pdfborder={0 0 0}]{hyperref}
\usepackage[square,comma,numbers,sort&compress]{natbib}
\usepackage{subcaption}
\usepackage{times}
\usepackage[T1]{fontenc}
\usepackage{relsize}

\usepackage{microtype}
\usepackage{iris}
\usepackage{pftools}
\usepackage{xfrac}
\usepackage{fontawesome}
\input{glyphtounicode}
\usepackage[normalem]{ulem}
\usepackage{multiaudience}
\usepackage{xcolor}

\SetNewAudience{conf}
\SetNewAudience{full}

\ifconfversion
\usepackage{xr}
\fi

\ifconfversion
\fi

\usepackage{pgfplots}
\pgfplotsset{width=10cm,compat=1.9}
\graphicspath{ {img/} }

\usepackage{tikz}
\usetikzlibrary{arrows,arrows.meta, shapes.geometric,decorations.pathreplacing,calc,positioning,matrix,tikzmark,intersections}
\usepackage{amsmath}
\usepackage{amssymb}
\usepackage{hyperref}
\usepackage{xspace}
\usepackage{booktabs}
\usepackage{multirow}
\usepackage{verbatim}
\usepackage{listings}
\usepackage{fancyvrb}
\usepackage[utf8]{inputenc}
\usepackage[group-separator={,}]{siunitx}
\usepackage[inline,final]{showlabels}
\usepackage{enumitem}

\def\Snospace~{\S{}}

\DeclareUnicodeCharacter{03C3}{$\sigma$}
\DeclareUnicodeCharacter{2264}{$\leq$}

\tikzstyle{every node}=[font=\small]
\setcitestyle{notesep={: }}

\conferenceinfo{SOSP '23}{October 23--26, 2023, Koblenz, Germany}
\CopyrightYear{2023}
\crdata{ISBN 979-8-4007-0229-7/23/10}
\newcommand{\doi}{10.1145/3600006.3613172}

\input{cmds}
\input{code/fmt.tex}

\begin{document}

\title{\Large \bf
  \sys{}: a Separation-Logic Library for Verifying Distributed Systems
  \showto{full}{\small (Extended Version)\footnote{This is the extended
  version of a paper appearing at SOSP 2023~\cite{sharma:grove};
  \extendedsymbol{} denotes sections not present in the conference
  version of the paper.}}
}

\author{
  Upamanyu Sharma \quad
  Ralf Jung$^\dag$ \quad
  Joseph Tassarotti$^\triangledown$ \quad
  M. Frans Kaashoek \quad
  Nickolai Zeldovich \\
  \em MIT CSAIL \qquad $^\dag$ ETH Zurich \qquad $^\triangledown$ New York University
}

\date{\vspace{-2\baselineskip}}

\maketitle

\begin{abstract}
\input{abstract}
\end{abstract}
\input{intro}

\input{system}

\input{design}

\input{case}

\input{modularity}

\input{impl}

\input{eval}

\input{related}

\input{conclusion}
\input{ack}

\bibliography{n-str,paper,n,n-conf}{}
\bibliographystyle{plainnat}

\end{document}

%% file: cmds.tex
\newcommand{\sys}{Grove\xspace}

\newif\ifdraft
\drafttrue 

\newcommand{\extendedsymbol}{\texorpdfstring{\fcolorbox{black}{black}{\textcolor{white}{\bf E}}}{[E]}}

\newcommand{\cc}[1]{\texorpdfstring{\mbox{\smaller[0.5]\texttt{\detokenize{#1}}}}{\texttt{\detokenize{#1}}}}

\newcommand{\loc}[1]{\num[group-separator={,},group-minimum-digits=4]{#1}}

\NewDocumentCommand \hoareCV {O{c} m m m m}{
  {\begin{aligned}[#1]
  &\curlybracket{#2} \\
  &\quad{#3} \\
  &\curlybracket{#4} \\
  &\curlybracket{#5}
  \end{aligned}}%
}

\renewcommand{\Ret}[1]{\text{\textbf{ret}}\,\, #1,\,}

\newcommand{\kvptsto}[2]{#1 \mapsto_{\text{kv}} #2}

\newcommand{\TLA}{$\text{TLA}^+$}

\newcommand{\textlogsm}[1]{\mbox{\smaller[0.5]\textsf{\upshape #1}}}
\newcommand{\FALSEsm}{\textlogsm{False}}

\newcommand{\tok}{\textlogsm{Tok}}

\newcommand{\aol}[2]{#1 \stackrel{\text{list}}{\mapsto} #2}
\newcommand{\aolb}[2]{#1 \stackrel{\text{list}}{\sqsupseteq} #2}
\newcommand{\aolro}[2]{#1 \stackrel{\text{list}}{\mapsto}_\square #2}

\newcommand{\fileptsto}[3]{\cc{``#1''} \stackrel{\text{file}}{\mapsto}_{#3} #2}

\newcommand{\currentepoch}[1]{\textlogsm{CurrentEpoch} \mapsto #1}
\newcommand{\currentepochlb}[1]{\textlogsm{CurrentEpoch} \geq #1}
\newcommand{\currenttime}[1]{\textlogsm{CurrentTime} \mapsto #1}

\newcommand{\lease}[2]{\ownGhost{}{#2}_{#1}}
\newcommand{\leaseexpiration}[2]{#1 \stackrel{\text{expires}}{\mapsto} #2}
\newcommand{\leaseexpirationlb}[2]{#1 \stackrel{\text{expires}}{\geq} #2}

\newcommand{\paxoswf}{\textlogsm{WF}}
\newcommand{\paxosstate}[1]{\textlogsm{Paxos} \mapsto #1}

\newcommand{\ownvsm}[1]{\textlogsm{VSM} \mapsto #1}

\newcommand{\kv}{vKV\xspace}

\makeatletter
\ifconfversion
\newcommand{\xref}[1]{\@ifundefined{r@#1}{\cite[\autoref*{full::#1}]{sharma:grove-extended}}{\autoref{#1}}}
\else
\newcommand{\xref}[1]{\autoref{#1}}
\fi
\makeatother

%% file: abstract.tex
\sys{} is a concurrent separation logic library for verifying
distributed systems.  \sys{} is the first to handle time-based leases,
including their interaction with reconfiguration, crash recovery,
thread-level concurrency, and unreliable networks.  This paper uses
\sys{} to verify several distributed system components written in Go,
including \kv{}, a realistic distributed multi-threaded key-value store.
\kv{} supports reconfiguration, primary/backup replication, and crash
recovery, and uses leases to execute read-only requests on any replica.
\kv{} achieves high performance (67-73\% of Redis on a single
core), scales with more cores and more backup replicas (achieving about
2$\times$ the throughput when going from 1 to 3 servers), and can safely
execute reads while reconfiguring.

%% file: intro.tex
\section{Introduction}

Large-scale applications run on many servers, and face a wide range of
challenges typical of distributed systems such as concurrency, crashes,
network outages, loosely-coupled clocks between servers, etc.  This means
that the developer has to consider a large number of subtle corner cases
and interactions, which in turn makes it difficult to ensure that the
application correctly handles all such cases.  Formal verification is an
attractive approach to rigorously establish correctness of such systems,
and in principle could help developers ensure that they correctly handle
all of the corner cases.

One particularly challenging and cross-cutting aspect of distributed
systems, which has not been addressed in prior work on verification, is
the use of leases.  Leases~\cite{gray:leases} are a widely used technique
in distributed systems. A lease is a promise that some aspect of the
system will not change for some duration of time (e.g., the primary
server will not be replaced for the next 5 seconds).
For instance, leases are used to ensure there is at most one Paxos leader trying
to run the replication protocol in Spanner~\cite{corbett:spanner};
and GFS~\cite{ghemawat:gfs}, Chubby~\cite{burrows:chubby}, and
DynamoDB~\cite{elhemali:dynamodb} have similar mechanisms.  Leases allow
a leader to execute read-only operations quickly, without having to
contact replicas to confirm that it is still the leader.  Leases are
challenging to use correctly because they interact with crash recovery
and reconfiguration (e.g., reconfiguration must wait until leases expire
before it can choose a new primary server) and node-local concurrency
(e.g., executing a read-only operation may require first checking that
a lease is valid, but in the time between the check and the operation
itself, the lease may have expired, and other threads may have executed
additional writes).

This paper presents \sys, a library based on concurrent separation
logic (CSL)~\citep{ohearn:csl} for reasoning about distributed systems,
where state is split between nodes, crashes discard a node's memory
state, network messages can be lost or duplicated, and nodes have loosely
synchronized clocks.  In CSL, a proof decomposes a system's state into
parts called resources that are logically owned by different threads.
Synchronization primitives, like mutexes, are used to transfer
ownership between threads.  \sys generalizes this notion of resources
and ownership to reason about distributed systems; in particular,
\sys{} introduces \emph{time-bounded invariants} to reason about
leases, extends Crash Hoare Logic~\cite{chen:fscq, chajed:perennial}
to reason about crashes in distributed systems, provides abstractions
for reasoning about append-only logs and monotonic epoch counters, and
provides a verified RPC library.  This
makes \sys{} the first to support verification of distributed systems
that use leases, including their interaction with crash recovery,
reconfiguration, concurrency, and unreliable networks.

To demonstrate \sys{}'s approach, we developed a number of distributed
system components written in Go (libraries, systems, and applications),
and specified and verified them using \sys{}.  As we explain in
\autoref{sec:design}, these components make extensive use of \sys{}'s
ownership and resources.  To name some examples: the proof of consistency
for a replicated log in a primary-backup replication library uses
ownership of logical append-only lists; the proof of crash recovery in
a durable storage library uses ownership of durable files; proofs about
RPCs that may re-execute many times use duplicable ownership (which can be
thought of as knowledge), since they cannot transfer ownership of unique
resources; and proofs about read operations that use leases
to avoid coordination in a state-machine replication library
uses time-bounded invariants to prove the state being read is not stale.

\input{components_table}

By using CSL, \sys{} enables modular reasoning: developers can verify
each component of a distributed system separately, and reason about
code line-by-line, rather than explicitly considering all possible
interleavings.  Nevertheless, these proofs still compose into a
complete proof of the entire distributed system.  For instance, our
case study builds a replicated key-value service (called \kv{}) out of
multiple independent components (RPC library, primary-backup replication,
state-machine replication, durable storage, configuration service, etc),
and builds an example bank application on top of \kv{} and a distributed
lock service.  The proof of the bank considers only the specifications
of the underlying \kv{} and lock service, and does not look at their
implementation.  At the same time, the composed proofs ensure there are
no subtle bugs due to surprising interactions between the components.
As we show in \autoref{sec:impl}, the proof-to-code ratio is about
12$\times$, on par with other distributed systems verification efforts,
which shows that handling leases, reconfiguration, concurrency, etc,
with \sys{} does not come at the cost of inflated proof effort.

\sys{}'s support for leases, concurrency, reconfiguration, and crash
recovery is crucial for verifying high-performance distributed systems.
\autoref{sec:eval} shows that \kv{} achieves good performance (67--73\%
of the throughput of Redis in a single-core unreplicated configuration),
scales well with the number of cores and the number of backup replicas
(going from 463,491 to 816,252 req/sec for a 95\%-read YCSB workload when
using 1 and 3 servers respectively), and is able to serve read requests
quickly and safely during reconfiguration.

To summarize, the contribution of this paper is \sys{}, which generalizes
concurrent separation logic (CSL) to support distributed systems with
RPCs, leases, replication, reconfiguration, and crash recovery.  The paper
provides lessons, insights, and techniques at several different levels.
For a general systems audience, \sys{} demonstrates that ownership-based
reasoning (using CSL) is valuable for distributed systems, by showing what
kinds of distributed systems can be verified, how verification catches
specific subtle bugs (the ``what if'' scenarios in \autoref{sec:case}),
and how CSL leads to modular development (\autoref{sec:modularity}).
For a verification audience, \sys{} presents techniques and ideas for
how to extend CSL to reason about distributed systems issues, such as
RPCs, leases, and replication, as we describe in \autoref{sec:design}.
These ideas may be helpful to researchers building frameworks for
verifying distributed systems.  Finally, the source code of \sys{} and its
case studies is publicly available,\footnote{Grove is available at
\url{https://github.com/mit-pdos/perennial} and the case studies are at
\url{https://github.com/mit-pdos/gokv}.} for experts that may want to adopt
\sys{}'s lower-level techniques for encoding distributed systems in the
Iris separation logic~\cite{jung:iris-1,krebbers:ipm,jung:iris-jfp}.

One limitation is that \sys{} is only able to verify safety
properties, ensuring that a system never returns the wrong
results.  \sys{} cannot verify liveness properties, such as ensuring
that the system will respond or otherwise make progress.

\input{components_graph}

%% file: components_table.tex
\begin{figure*}[t]
\small
\centering
\begin{tabular}{lp{0.7in}p{1.3in}l}
{\bf Component} & {\bf Code} & {\bf Spec and proof} & {\bf Description} \\
  \toprule

\cc{bank} & \xref{sec:system:bank} & \xref{sec:case:bank} & Uses \kv{} and lock service to execute bank transactions \\
\cc{lockservice} & \xref{sec:system:lockservice} & \xref{sec:case:bank} & Distributed lock service implemented on top of \kv{} \\
\cc{cachekv} & \xref{sec:system:cachekv} & \xref{sec:case:cache} & Uses leases for linearizable key-value caching on client \\
\kv{} & \xref{sec:system:kv} & \xref{sec:modularity:top} & Handles key-value state and operations (\cc{Get}, \cc{Put}, \cc{CondPut}) \\
\cc{exactlyonce} & \xref{sec:system:exactlyonce} & \xref{sec:design:exactlyonce}, \xref{sec:design:prophecy} & Tracks and handles duplicate operations for exactly-once semantics \\
\cc{clerk} & \xref{sec:system:clerk} & \xref{sec:modularity:top} & Issues operations to replicated state machine \\
\cc{storage} & \xref{sec:system:storage} & \xref{sec:design:chl} & Stores application state and handles state transfers \\
\cc{configservice} & \xref{sec:system:reconfig} & \xref{sec:case:configpaxos}, \xref{sec:modularity:configpaxos} & Tracks the changing set of replica servers and issues epoch leases \\
\cc{paxos} & \xref{sec:system:configpaxos} & \xref{sec:case:configpaxos} & Simple Paxos-based fault-tolerant replication for \cc{configservice} \\
\cc{reconfig} & \xref{sec:system:reconfig} & \xref{sec:case:reconfig} & Adds or removes replicas, using the config service \\
\cc{replica} & \xref{sec:system:pb}, \xref{sec:system:leasereads} &
  \xref{sec:design:ghost}, \xref{sec:design:invariants}, \xref{sec:design:rpc}, \newline
  \xref{sec:case:pb}, \xref{sec:case:reconfig}, \xref{sec:case:leases}, \xref{sec:case:ro} & Stores and replicates operations between primary and backups \\
Lease abstraction & --- & \xref{sec:design:leases}, \xref{sec:design:clocks} & Time-bounded invariant abstraction to reason about leases \\
\cc{rpc} & \xref{sec:system:rpc} & \xref{sec:design:rpc} & Unreliable request/response communication \\

\end{tabular}
\caption{Components verified using \sys{} as case studies.
  \begin{shownto}{conf} Some components are presented only in the extended version of this paper~\cite{sharma:grove-extended}.\end{shownto}
}
\label{fig:components}
\end{figure*}

%% file: components_graph.tex
\begin{figure}[ht]
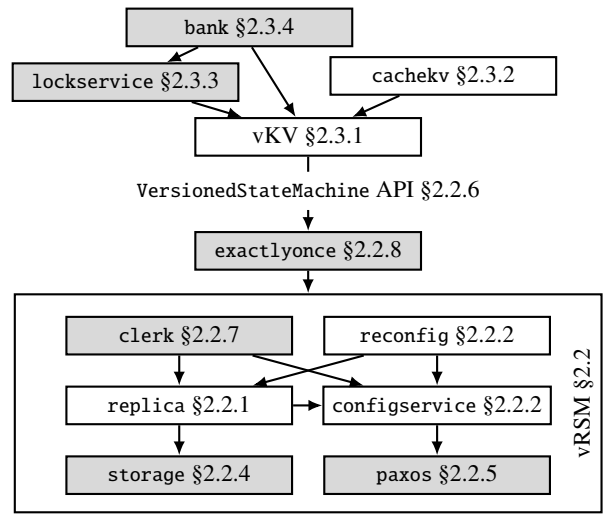

  \centering
\include{img/components.tex}
\vspace{-2\baselineskip}
\caption{Case study components.  An arrow $A \to B$ means $A$ uses $B$.
Gray components are described only in the extended version of
this paper\begin{shownto}{conf}~\cite{sharma:grove-extended}\end{shownto}.}
\label{fig:componentgraph}
\end{figure}

%% file: img/components.tex
\begin{tikzpicture}[>=latex]
\tikzset{
    box/.style = {
        rectangle,
        draw           = black, thick,
        minimum width  = 3.0cm,
        minimum height = 0.5cm,
        align=center,
    },
    ->/.style = {thick, arrows={-Latex[angle=45:2mm]}
    },
}
\newlength{\margin}
\setlength{\margin}{4mm}

\node[box, fill=gray!30, anchor=north,]
  (exactlyonce) at (0,0) {\cc{exactlyonce} \xref{sec:system:exactlyonce}};

\node[box, anchor=north east,
  fill=gray!30,
  below left=1.5\margin and \margin/2 of exactlyonce.south,
] (clerk) {\cc{clerk} \xref{sec:system:clerk}};

\node[box, anchor=west, right=\margin of clerk.east]
  (reconfig) {\cc{reconfig} \xref{sec:system:reconfig}};

\node[box, anchor=north, below=\margin of clerk.south]
  (replica) {\cc{replica} \xref{sec:system:pb}};

\node[box, anchor=north, below=\margin of reconfig.south]
  (configservice) {\cc{configservice} \xref{sec:system:reconfig}};

\node[box, anchor=north, below=\margin of configservice.south, fill=gray!30]
  (paxos) {\cc{paxos} \xref{sec:system:configpaxos}};

\node[box, fill=gray!30, anchor=north, below=\margin of replica.south]
  (storage) {\cc{storage} \xref{sec:system:storage}};

\node[anchor=south, above=0.7\margin of exactlyonce.north]
  (vsm) {\cc{VersionedStateMachine} API \xref{sec:system:vsm}};

\node[box, anchor=south, above=0.5\margin of vsm.north]
  (kv) {\kv{} \xref{sec:system:kv}};

\node[box, fill=gray!30, anchor=south east,
  above left=\margin/2 and -1.5\margin of kv.north west,]
  (lockservice) {\cc{lockservice} \xref{sec:system:lockservice}};

\node[box, fill=gray!30, anchor=south, above=0.5\margin of lockservice.north east]
  (bank) {\cc{bank} \xref{sec:system:bank}};

\node[box, anchor=west, above right =\margin of kv.north]
  (cachekv) {\cc{cachekv} \xref{sec:system:cachekv}};

\node[rectangle, draw, thick, anchor=north,
  minimum height=2.9cm, minimum width=7.8cm,
  below=0.75\margin of exactlyonce,
  align=right,
] (vrsm) {};

\node[anchor=center, above=0 of vrsm.east, rotate=90] (vrsm_text) {vRSM \xref{sec:system:rsm}};

\draw[->] (clerk) -- (replica);
\draw[->] (clerk) -- (configservice);
\draw[->] (reconfig) -- (replica);
\draw[->] (reconfig) -- (configservice);
\draw[->] (configservice) -- (paxos);

\draw[->] (replica) -- (configservice);
\draw[->] (replica) -- (storage);

\draw[->] (exactlyonce) -- (vrsm);
\draw[->] (kv) -- (vsm) -- (exactlyonce);
\draw[->] (lockservice) -- (kv);
\draw[->] (bank) -- (lockservice);
\draw[->] (bank) -- (kv);
\draw[->] (cachekv) -- (kv);

 \end{tikzpicture}

%% file: system.tex
\section{Motivating case studies}
\label{sec:system}

\sys{}'s goal is to enable verification of distributed system components
in a way that allows composing them into a single proof
for the entire system.  To illustrate the verification
challenges that \sys{} aims to address, this section presents a number of
components typically seen in distributed systems, shown in \autoref{fig:components},
spanning from RPC and
storage at the lowest level, to libraries for replicated state machines,
key-value stores, and locking, to application-level
code such as a bank example.  Distributed systems challenges, such as
concurrency, crashes, clocks, etc, show up in many of these components,
and a key benefit of Grove is that it provides a consistent framework
for handling these issues in the specifications and proofs of each
component, which in turn allows combining these components into larger
verified systems.  The components fit together to build \kv{}, a
replicated key-value store, as well as applications on top of it, as
shown in \autoref{fig:componentgraph}.

These components use sophisticated techniques to achieve high
performance and strong correctness guarantees.  For instance, they
use threads on each machine to execute RPCs in parallel; store data
durably on disk (using a separate thread for performance) and recover
their state after a crash; batch disk writes and pipelines requests for
improved performance; achieve linearizability even in the presence of
retransmission, crashes, and in-flight client requests while adding or
removing servers through reconfiguration; and use leases to coordinate
the execution of read-only requests at each replica with reconfiguration.

\subsection{RPC library}
\label{sec:system:rpc}

An important building block for distributed systems is RPC, which allows a
client to invoke a procedure on a remote server.  For instance, a client
invocation of \cc{rpcClient.Call(``f'', args)} invokes \cc{f(args)}
on the server to which \cc{rpcClient} is connected.  The \cc{rpc}
library provides \emph{unreliable} RPCs, meaning that one invocation by a client
can result in the server running the corresponding function one, zero,
or many times. This is because the underlying network may drop, reorder,
or duplicate packets. Applications typically do not directly invoke RPCs;
rather, applications use various \emph{clerks}, which wrap RPCs with
additional handling (such as adding request IDs, retrying, etc).

\subsection{Replicated state machine library}
\label{sec:system:rsm}

\begin{figure}[ht]
  \centering
\includegraphics[width=\columnwidth]{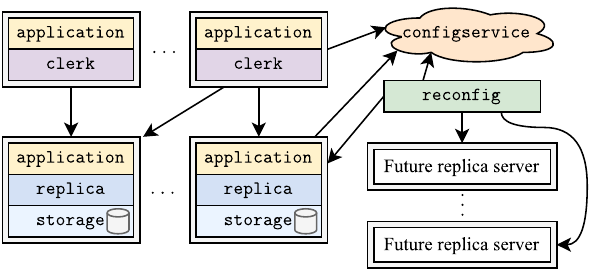}
\vspace{-1.5\baselineskip}
\caption{A running vRSM system.  Double borders represent machines.
  Arrows represent RPCs.  The cloud represents the replicated
  configuration service.  \cc{reconfig} represents an operator
  performing reconfiguration.}
\label{fig:system}
\end{figure}

The focal point of our case study is a replicated state machine library
called vRSM, as shown in \autoref{fig:system}.  vRSM replicates a state
machine supplied by the application (the exact interface is shown in
\showto{conf}{the extended version of this paper~}\xref{fig:vsm_code}).
\xref{sec:system:apps} discusses how applications use this interface.
vRSM is implemented in several components, which this subsection describes.
The components each handle a different aspect of state machine replication,
allowing, for instance, durability to be implemented separately from the
replication protocol.

\subsubsection{\cc{replica} server: replicating writes}
\label{sec:system:pb}

The \cc{replica} component manages copies of the state machine being replicated.
A \cc{replica} server is either a primary and handles write requests from
clients, or else is a backup (we discuss the handling of reads later in
\autoref{sec:system:leasereads}). Upon receiving an operation, a primary server
applies it locally and then replicates it to all backup servers before replying
to the client, as shown in \autoref{fig:apply}.  To replicate an operation, the
primary spawns threads to send RPCs concurrently to each backup and then waits
for all the threads to finish (using a Go \cc{WaitGroup}) to know that the
operation is committed (i.e.  applied by all \cc{replica} servers). Backup
replicas handle these RPCs by also applying the operation locally.
\cc{s.stateLogger} takes care of managing the RSM state, as we describe in
\xref{sec:system:storage}.

\begin{figure}[ht]
\include{code/primary_apply}
\vspace{-3\baselineskip}
\caption{Simplified primary server code, with error handling omitted.}
\label{fig:apply}
\end{figure}

This protocol requires the primary to replicate the operation to \emph{all}
servers before replying to a client, so if even a single backup is unavailable,
the replication protocol is blocked. To unblock the system, an operator or an
automatic failure detector can remove unresponsive servers (and add new ones) by
invoking the \cc{reconfig} component, described next.

\subsubsection{\cc{reconfig} using \cc{configservice}}
\label{sec:system:reconfig}
The \cc{reconfig} component allows adding or removing \cc{replica} servers by
making use of sequentially numbered \emph{epochs} and a \cc{configservice}
component. An epoch typically corresponds to a configuration---that is, a set of
servers with one designated as the primary.  We call such epochs \emph{live},
even if such an epoch has been superseded by another one.  However, some epochs
may not have a corresponding configuration, if that epoch never started running
(e.g. because a node running \cc{reconfig} crashed); we call such epochs \emph{reserved}.
The \cc{configservice}
keeps track of the latest epoch number and the most recent configuration (a list
of server addresses), which may be from an earlier epoch if the
current epoch is not live.

Clients can invoke operations concurrently with reconfiguration, which runs the
risk of a client's operations being applied in an old configuration after the
new configuration has already started, thereby missing these operations in the
new config.  To prevent this, reconfiguration first \emph{seals} one of the
servers from the old epoch.  A sealed server no longer modifies its state until
it enters a new epoch, at which point it becomes unsealed. Sealed servers may
still handle read requests. Sealing allows the reconfiguration process to get a
stable checkpoint of the system state and ensure all of the servers in the new
configuration have consistent state before entering the new epoch.

\begin{figure}[ht]
\include{code/config}
\vspace{-3\baselineskip}
\caption{Interface provided by the configuration service clerk.}
\label{fig:config_code}
\end{figure}

\begin{figure}[ht]
\include{code/reconfig}
\vspace{-3\baselineskip}
\caption{Simplified \cc{reconfig} code, with most error handling omitted.}
\label{fig:reconfig_code}
\end{figure}

Reconfiguration involves coordination between the configuration service,
the old servers, and the new servers.  \autoref{fig:config_code} shows
the API for the configuration service, and \autoref{fig:reconfig_code}
shows the code for reconfiguration, invoked to change to a new set of
servers specified by the \cc{newServers} argument. Not shown is the
monitoring logic that decides when to call this function or which new
servers to choose; correctness (safety) is independent of that logic,
and Grove does not prove liveness.
Reconfiguration consists of the following steps:

\begin{CompactEnumerate}
\item Ask the configuration service to atomically create a new epoch
  and return the new epoch's number as well as the latest previous configuration
  (line 2).

\item Seal a replica server from the previous configuration and fetch
  its key-value mappings (line 6).

\item Initialize state on all new servers with the state from the old replica,
  informing them of the new epoch (line 19).

\item Make the new epoch live at the configuration service, by sending
  it the new configuration (line 26).

\item Enable the primary in the new configuration, which allows the
  primary to start processing write requests
  (line 29).

\end{CompactEnumerate}

In case of a network partition, it is possible that both sides of the
partition will try to initiate reconfiguration. One might worry that
this would lead to two copies of the system with diverging states.
This possibility is ruled out with the help of the
configuration service, which accepts only the highest-numbered new
epoch in its \cc{WriteConfig} RPC handler, together with the replicas'
\cc{SetNewEpochState} handler, which rejects state from lower-numbered epochs.
As a result, the reconfiguration process that has the higher new epoch
from \cc{GetNewEpochAndConfig()} will win.

\subsubsection{\cc{replica} server: lease-based reads}
\label{sec:system:leasereads}

Any \cc{replica} server (primary as well as backup) can serve linearizable reads
without communicating with other servers by using leases, as shown in
\autoref{fig:lease_code}.  Leases avoid the possibility of one server returning
stale reads if reconfiguration happens and the new configuration has executed
additional writes not seen by this server.  Specifically, every replica runs a
background thread that contacts the configuration service to obtain or extend a
lease that promises the configuration service will not change the current epoch
number (and thus not reconfigure) until lease expiration (e.g., 1
second from the time the lease is issued). All servers can serve read requests
because this lease is a promise about the epoch number, rather than anything
specific to a particular server's state.

When a \cc{replica} server receives a read-only operation, and its lease is
still valid, it computes the response from its local state. The \cc{replica}'s
local state includes all committed operations since committed operations must be
acknowledged by all servers.  However, the state may also include ongoing write
operations that have not yet been committed.  To ensure that the client's
observed read does not roll back due to a crash or reconfiguration, the replica
waits for all the previous writes that the read depends on to be committed
before sending the result to the client. As part of executing the read
operation, \cc{LocalRead}'s job is to determine which prior requests the read
depends on, returning the appropriate \cc{idx} value as shown in
\autoref{fig:lease_code}; it is always safe to return \cc{idx = s.nextIndex}.
If reconfiguration happens during \cc{waitForCommitted}, the server tells the
client to retry.

\begin{figure}[ht]
\include{code/lease}
\vspace{-3\baselineskip}
\caption{Simplified code for handling read-only operations.}
\label{fig:lease_code}
\end{figure}

Since the clocks on different nodes might be slightly out of sync with
each other, \sys{} provides a TrueTime-like API~\cite{corbett:spanner}
for accessing the current time, \cc{GetTimeRange()}.  This function
returns a pair of timestamps, \cc{earliest} and \cc{latest}, which
provide lower and upper bounds for the current time.

\begin{Subsubsection}{full}{\cc{storage} library for replicas \extendedsymbol{}}
\label{sec:system:storage}

Replica servers manage durable state with a \cc{storage} library that provides a
``state logger'' to durably log new operations in an append-only file.  To get
good performance, the state logger buffers appends in memory while a background
thread asynchronously appends and syncs the buffer to the file.  The library
provides a \cc{Wait()} function that allows waiting until a prefix of the file
has been made durable.  The \cc{replica} library uses \cc{Wait()} to ensure
changes are durably stored before replying to an RPC (not shown in our
simplified code examples).
\end{Subsubsection}

\begin{Subsubsection}{full}{Fault-tolerant \cc{configservice} using \cc{paxos} \extendedsymbol{}}
\label{sec:system:configpaxos}

To handle server failures for the configuration service itself,
\cc{configservice} relies on a simple Paxos-based replication library
called \cc{paxos}.  \cc{paxos} operates on a fixed set of servers (new
servers cannot be added at runtime), but requires only a majority of
the servers to process requests.  \cc{paxos} uses a leader to coordinate
operations, but allows changing the leader if the previous one seems to
have crashed.

The code structure of \cc{paxos} largely follows the primary-backup
replication library, with a few key differences.  First, instead of
relying on an external config service to choose new epoch numbers,
\cc{paxos} chooses new epoch numbers on its own, which it can do safely
because the set of servers does not change.  Second, instead of requiring
every server to commit an operation, \cc{paxos} requires only a majority,
which also means that a new leader must obtain the latest state from a
majority of other servers, rather than just one.  Third, \cc{paxos}
is much simpler than the primary-backup replication library: it does not
use leases, and it sends and writes the entire state to disk on every update
rather than appending operations to a log. This means that \cc{paxos} has lower
performance for write operations, which is acceptable for the configuration
service. Finally, \cc{paxos} provides fast but weakly consistent reads.

The interface provided by \cc{paxos} is shown in \autoref{fig:paxos_code}.
\cc{WeakRead} returns the entire current state replicated by \cc{paxos},
as stored on the server where \cc{WeakRead} is invoked.  The resulting
state might be stale (if other servers have committed new operations in
the meantime) or the resulting state might not even be committed (if this
state was never acknowledged by a majority of servers).  \cc{configservice}
implements \cc{GetConfig} using \cc{WeakRead}, despite its weak semantics,
because the caller of \cc{GetConfig} is the vRSM clerk, which can handle
stale or even incorrect results, and because using \cc{WeakRead} ensures
that \cc{GetConfig} is fast.

\begin{figure}[ht]
\include{code/paxos}
\vspace{-3\baselineskip}
\caption{Interface provided by \cc{paxos}.}
\label{fig:paxos_code}
\end{figure}

To execute write operations, such as \cc{ReserveEpochAndGetConfig}
as shown in \autoref{fig:configimpl_code}, \cc{configservice} uses the
\cc{Begin} method (which should be invoked on the leader, otherwise the
operation will be unable to commit).  \cc{Begin} returns the current
replicated state from the local server, as well as a callback function
\cc{commit} that \cc{configservice} will use to try to commit its new
state.  The \cc{commit}
callback is the core of \cc{paxos}: it actually talks to other servers,
unlike \cc{WeakRead} and \cc{Begin}.  It tries to replicate the new state
to a majority of servers, while checking that no other operations have
been committed in the meantime.  \cc{commit} could succeed in getting
a quorum of servers to accept this new state, or it might fail because
another leader has been chosen.

\begin{figure}[ht]
\include{code/configimpl}
\vspace{-3\baselineskip}
\caption{Implementation of the \cc{ReserveEpochAndGetConfig} RPC handler
  in \cc{configservice}, built on top of \cc{paxos}.  This handler is
  invoked by the \cc{configservice} clerk shown in \autoref{fig:config_code}.}
\label{fig:configimpl_code}
\end{figure}

Finally, if the current leader appears to have crashed, the operator
could call \cc{TryBecomingLeader} to choose another server as the leader;
it is always safe to call \cc{TryBecomingLeader}.

\end{Subsubsection}

\begin{Subsubsection}{full}{Versioned state machine API \extendedsymbol{}}
\label{sec:system:vsm}

To build an application on top of vRSM, the developer must implement the
\emph{versioned state machine} interface shown in \autoref{fig:vsm_code}.  The
\cc{Apply()} executes a application-level read/write operation and \cc{Read()}
executes a application-level read operation against the against the current
in-memory state, while \cc{SetState()} and \cc{GetState()} allow serializing the
in-memory state.  Using this developer-provided interface, the vRSM library
takes care of checkpointing the state on disk and copying the state to new
replicas.  For example, the call to \cc{GetStateAndSeal()} in
\autoref{fig:reconfig_code} ultimately uses the developer-provided
\cc{GetState()} method to checkpoint the current state, and the call to
\cc{SetNewEpochState()} in \autoref{fig:reconfig_code} ultimately calls the
developer-provided \cc{SetState()} method to initialize the replica's local
state.

\begin{figure}[ht]
\include{code/vsm}
\vspace{-3\baselineskip}
\caption{Interface for a versioned state machine.}
\label{fig:vsm_code}
\end{figure}

As described earlier, one complication is that read operations may observe
writes that have not been replicated to all servers yet and thus have to wait
for those operations to be committed.  As an optimization, the
application-provided \cc{Read()} method can specify which writes the read result
depends on, by returning the \cc{idx} of the most recent write dependency as
the first part of its return value. This allows reads to return quickly, without
waiting for commits of recent writes, if the result is not reading a recent
write.  This propagates into the \cc{idx} value returned by
\cc{s.stateLogger.LocalRead()} in \autoref{fig:lease_code} and determines what
committed operations the \cc{replica} library waits for.

\end{Subsubsection}

\begin{Subsubsection}{full}{vRSM \cc{clerk} \extendedsymbol{}}
\label{sec:system:clerk}

vRSM provides a client \cc{clerk} library, which hides the complexity of issuing
requests to vRSM over the network.  The clerk API is shown in
\autoref{fig:rsmclerk_code}. Applications can use the clerk library on many
client machines to access vRSM. To execute an operation, the clerk needs to
know the address of the \cc{replica} servers. The clerk initially obtains this
information from the configuration service and caches it locally. Calling
\cc{clerk.Apply(op)} issues read-write operations to the primary while
\cc{clerk.Read(op)} issues a read-only operation to any replica.  If a server
indicates that it is no longer the primary (or replica, for read-only
operations), the clerk asks the config server for the new server information and
retries. Because of retries, it is possible for a single \cc{clerk.Apply()}
call to result in an operation being applied more than once. A higher-level
library handles deduplicating operations (\autoref{sec:system:exactlyonce}).

\begin{figure}[ht]
\include{code/rsmclerk}
\vspace{-3\baselineskip}
\caption{Interface provided by the vRSM client clerk.}
\label{fig:rsmclerk_code}
\end{figure}

\end{Subsubsection}

\begin{Subsubsection}{full}{\cc{exactlyonce} library \extendedsymbol{}}
\label{sec:system:exactlyonce}

The \cc{exactlyonce} library helps applications using vRSM ensure that
operations execute exactly-once.  It consists of a new clerk that wraps over the
vRSM clerk (which potentially duplicates operations through retries) and a state
machine transformer that adds duplicate detection and handling to an
application-level state machine.  The clerk API is the same as the underlying
vRSM clerk, but additionally guarantees that operations are not applied more
than once.  To achieve this, the clerk adds a unique request ID made up of a
client ID and sequence number to each operation. On the other side, the
\cc{exactlyonce} state machine transformer augments an application-level state
machine with a reply table to keep track of previously applied requests along
with their replies. Upon getting a new request, the \cc{exactlyonce} library
calls the application state machine's \cc{Apply} function and stores the reply
in the reply table. Upon getting a duplicate, the library does not call into the
application state machine and instead returns with the previous reply.
For read-only operations, \cc{exactlyonce} ignores the reply table and calls
\cc{Read} on the application state machine.

\end{Subsubsection}

\subsection{Applications on top of vRSM}
\label{sec:system:apps}

\subsubsection{\kv{}}
\label{sec:system:kv}

\kv{} is implemented on top of vRSM and the \cc{exactlyonce} library. The
server-side part of \kv{} is an implementation of the state machine interface
expected by vRSM.  The client-side part of \kv{} is a clerk implemented on top
of the \cc{exactlyonce} clerk, with the API shown in \autoref{fig:kvclerk_code}.
By building on top vRSM, the implementation of \kv{} itself is simple: it
consists of (de)serialization methods to turn key-value operations into byte
slices and a few functions to read and update an in-memory map.
In addition to storing a map of keys to values, \kv{} also stores a map
from keys to the index of the last operation that modified that key,
which allows \kv{} to take advantage of vRSM's versioned state machine
interface (\xref{sec:system:vsm}) to improve the performance of reads.

\begin{figure}[ht]
\include{code/kvclerk}
\vspace{-3\baselineskip}
\caption{Interface provided by the \kv{} client clerk.}
\label{fig:kvclerk_code}
\end{figure}

\subsubsection{Lease-based client-side caching}
\label{sec:system:cachekv}

As another example of using leases, \cc{cachekv} is a lease-based client-side
caching library that works by storing both data and lease expiration times in
\kv{}.
\autoref{fig:leasekv_code} shows \cc{GetAndCache} function, which returns the
value of the specified \cc{key} and caches it internally for \cc{cachetime}
time. It uses \cc{CondPut} to atomically increase the lease duration, which
ensures that a concurrent modification did not change the value since the
\cc{Get} on line 4. Similarly, \cc{CacheKv}'s \cc{Put} function (not shown) uses
\cc{CondPut} to ensure the value is only changed if the lease is expired.
Finally, \cc{CacheKv}'s \cc{Get} function first tries reading from \cc{k.cache}
and only invokes the \cc{Get} on \kv{} if the value is not cached. The
client-caching library is simple, but exemplifies how leases can be used for
cache consistency~\cite{gray:leases}.

\begin{figure}[ht]
\include{code/leasekv}
\vspace{-3\baselineskip}
\caption{Simplified code for getting a lease on a key and caching it.}
\label{fig:leasekv_code}
\end{figure}

\begin{Subsubsection}{full}{Lock service  \extendedsymbol{}}
\label{sec:system:lockservice}

\autoref{fig:lockservice_code} shows the interface provided by the
lock service, built on top of \kv{}.  The lock service uses \kv{}'s
conditional-put \cc{CondPut()} operation to implement locks, with one
lock corresponding to one key-value pair.  The lock service provides
a specification for its \cc{Acquire()} and \cc{Release()} methods that
allows applications to implement exclusive locking, such as accounts in
our bank example, in the style of a traditional concurrent separation
logic lock specification~\cite{ohearn:csl}.  This specification is quite
different from the specifications of the underlying \kv{} methods like
\cc{CondPut()}, and makes it easy for the bank example to keep separation
logic resources protected by locks that the lock service provides.

\begin{figure}[ht]
\include{code/lockservice}
\vspace{-3\baselineskip}
\caption{Interface provided by the lock service.}
\label{fig:lockservice_code}
\end{figure}

\end{Subsubsection}

\begin{Subsubsection}{full}{Bank transactions  \extendedsymbol{}}
\label{sec:system:bank}

As a top-level application, we implemented a toy bank application, which
uses transactions built on top of the \kv{} clerk and
lock service interfaces, and does not depend on the details of how those
interfaces are implemented.  However, the specifications for the \kv{}
clerk and lock service are strong enough to prove correctness for the
bank's transactions.

The bank uses an instance of \kv{} to store its account state, with one
key-value pair used to store the balance of one account.  The bank uses
the lock service (with its own separate instance of \kv{}) to handle
concurrent access to accounts.  Every \cc{Transfer(src, dst, amt)}
operation obtains two locks, on the \cc{src} and \cc{dst} accounts (sorted
to avoid deadlock) before accessing their respective account balances
in \kv{}.  This ensures that concurrent transfers are safe to execute,
and allows for concurrency when transfers access different accounts.
The \cc{Audit()} function grabs locks for all accounts, computes the
total balance by retrieving each account's balance, and then releases
all of the locks.

If one of the bank nodes crashes, the locks held by any threads on that
node in the lock service will remain locked.  Recovering from this would
require some form of undo or redo logging; for instance, the bank threads
could send undo log entries to the lock service.  We have not implemented
this in the bank prototype.

\end{Subsubsection}

%% file: code/primary_apply.tex
\begin{Verbatim}[commandchars=\\\{\},numbers=left,firstnumber=1,stepnumber=1,codes={\catcode`\$=3\catcode`\^=7\catcode`\_=8\relax},fontsize=\scriptsize,numbersep=6pt,xleftmargin=0.2in]
\PY{k+kd}{func} \PY{p}{(}\PY{n+nx}{s} \PY{o}{*}\PY{n+nx}{PrimaryServer}\PY{p}{)} \PY{n+nx}{Apply}\PY{p}{(}\PY{n+nx}{op}\PY{p}{)} \PY{n+nx}{Result} \PY{p}{\PYZob{}}
  \PY{n+nx}{s}\PY{p}{.}\PY{n+nx}{mutex}\PY{p}{.}\PY{n+nx}{Lock}\PY{p}{(}\PY{p}{)}
  \PY{n+nx}{nextIndex} \PY{o}{:=} \PY{n+nx}{s}\PY{p}{.}\PY{n+nx}{nextIndex}
  \PY{n+nx}{e} \PY{o}{:=} \PY{n+nx}{s}\PY{p}{.}\PY{n+nx}{epoch}
  \PY{n+nx}{s}\PY{p}{.}\PY{n+nx}{nextIndex} \PY{o}{+=} \PY{l+m+mi}{1}
  \PY{n+nx}{res} \PY{o}{:=} \PY{n+nx}{s}\PY{p}{.}\PY{n+nx}{stateLogger}\PY{p}{.}\PY{n+nx}{LocalApply}\PY{p}{(}\PY{n+nx}{op}\PY{p}{)}
  \PY{n+nx}{s}\PY{p}{.}\PY{n+nx}{mutex}\PY{p}{.}\PY{n+nx}{Unlock}\PY{p}{(}\PY{p}{)}

  \PY{n+nx}{wg} \PY{o}{:=} \PY{n+nb}{new}\PY{p}{(}\PY{n+nx}{WaitGroup}\PY{p}{)}
  \PY{k}{for} \PY{n+nx}{j} \PY{o}{:=} \PY{l+m+mi}{0}\PY{p}{;} \PY{n+nx}{j} \PY{p}{\PYZlt{}} \PY{n+nb}{len}\PY{p}{(}\PY{n+nx}{s}\PY{p}{.}\PY{n+nx}{backupClerks}\PY{p}{)}\PY{p}{;} \PY{n+nx}{j}\PY{o}{++} \PY{p}{\PYZob{}}
    \PY{n+nx}{wg}\PY{p}{.}\PY{n+nx}{Add}\PY{p}{(}\PY{l+m+mi}{1}\PY{p}{)}
    \PY{k}{go} \PY{k+kd}{func}\PY{p}{(}\PY{n+nx}{j} \PY{k+kt}{int}\PY{p}{)} \PY{p}{\PYZob{}}
      \PY{n+nx}{s}\PY{p}{.}\PY{n+nx}{backupClerks}\PY{p}{[}\PY{n+nx}{j}\PY{p}{]}\PY{p}{.}\PY{n+nx}{ApplyAsBackupRPC}\PY{p}{(}\PY{n+nx}{e}\PY{p}{,} \PY{n+nx}{nextIndex}\PY{p}{,} \PY{n+nx}{op}\PY{p}{)}
      \PY{n+nx}{wg}\PY{p}{.}\PY{n+nx}{Done}\PY{p}{(}\PY{p}{)}
    \PY{p}{\PYZcb{}} \PY{p}{(}\PY{n+nx}{j}\PY{p}{)}
  \PY{p}{\PYZcb{}}

  \PY{n+nx}{wg}\PY{p}{.}\PY{n+nx}{Wait}\PY{p}{(}\PY{p}{)}
  \PY{k}{return} \PY{n+nx}{res}
\PY{p}{\PYZcb{}}
\end{Verbatim}

%% file: code/config.tex
\begin{Verbatim}[commandchars=\\\{\},numbers=left,firstnumber=1,stepnumber=1,codes={\catcode`\$=3\catcode`\^=7\catcode`\_=8\relax},fontsize=\scriptsize,numbersep=6pt,xleftmargin=0.2in]
\PY{c+c1}{// Reserve a new epoch number for reconfiguration, and}
\PY{c+c1}{// return the current configuration (set of servers).}
\PY{k+kd}{func}\PY{+w}{ }\PY{p}{(}\PY{n+nx}{ck}\PY{+w}{ }\PY{o}{*}\PY{n+nx}{Clerk}\PY{p}{)}\PY{+w}{ }\PY{n+nx}{ReserveEpochAndGetConfig}\PY{p}{(}\PY{p}{)}\PY{+w}{ }\PY{p}{(}\PY{k+kt}{uint64}\PY{p}{,}\PY{+w}{ }\PY{p}{[}\PY{p}{]}\PY{n+nx}{Address}\PY{p}{)}

\PY{c+c1}{// Return current configuration, used by clients to}
\PY{c+c1}{// determine what servers to talk to.}
\PY{k+kd}{func}\PY{+w}{ }\PY{p}{(}\PY{n+nx}{ck}\PY{+w}{ }\PY{o}{*}\PY{n+nx}{Clerk}\PY{p}{)}\PY{+w}{ }\PY{n+nx}{GetConfig}\PY{p}{(}\PY{p}{)}\PY{+w}{ }\PY{p}{[}\PY{p}{]}\PY{n+nx}{Address}

\PY{c+c1}{// Set new configuration, making epoch live, as long as no}
\PY{c+c1}{// higher\PYZhy{}numbered epoch has been reserved.}
\PY{k+kd}{func}\PY{+w}{ }\PY{p}{(}\PY{n+nx}{ck}\PY{+w}{ }\PY{o}{*}\PY{n+nx}{Clerk}\PY{p}{)}\PY{+w}{ }\PY{n+nx}{TryWriteConfig}\PY{p}{(}\PY{n+nx}{epoch}\PY{+w}{ }\PY{k+kt}{uint64}\PY{p}{,}
\PY{+w}{                                }\PY{n+nx}{config}\PY{+w}{ }\PY{p}{[}\PY{p}{]}\PY{n+nx}{Address}\PY{p}{)}\PY{+w}{ }\PY{n+nx}{Error}

\PY{c+c1}{// Get a lease for specified epoch, as long as it\PYZsq{}s the current}
\PY{c+c1}{// epoch, returning the new lease expiration time.}
\PY{k+kd}{func}\PY{+w}{ }\PY{p}{(}\PY{n+nx}{ck}\PY{+w}{ }\PY{o}{*}\PY{n+nx}{Clerk}\PY{p}{)}\PY{+w}{ }\PY{n+nx}{GetLease}\PY{p}{(}\PY{n+nx}{epoch}\PY{+w}{ }\PY{k+kt}{uint64}\PY{p}{)}\PY{+w}{ }\PY{p}{(}\PY{n+nx}{Error}\PY{p}{,}\PY{+w}{ }\PY{k+kt}{uint64}\PY{p}{)}
\end{Verbatim}

%% file: code/reconfig.tex
\begin{Verbatim}[commandchars=\\\{\},numbers=left,firstnumber=1,stepnumber=1,codes={\catcode`\$=3\catcode`\^=7\catcode`\_=8\relax},fontsize=\scriptsize,numbersep=6pt,xleftmargin=0.2in]
\PY{k+kd}{func}\PY{+w}{ }\PY{n+nx}{Reconfigure}\PY{p}{(}\PY{n+nx}{newServers}\PY{+w}{ }\PY{p}{[}\PY{p}{]}\PY{n+nx}{Address}\PY{p}{)}\PY{+w}{ }\PY{p}{\PYZob{}}
\PY{+w}{  }\PY{n+nx}{newEpoch}\PY{p}{,}\PY{+w}{ }\PY{n+nx}{oldServers}\PY{+w}{ }\PY{o}{:=}\PY{+w}{ }\PY{n+nx}{configClerk}\PY{p}{.}\PY{n+nx}{ReserveEpochAndGetConfig}\PY{p}{(}\PY{p}{)}

\PY{+w}{  }\PY{c+c1}{// get state from a server from old config}
\PY{+w}{  }\PY{n+nx}{oldClerk}\PY{+w}{ }\PY{o}{:=}\PY{+w}{ }\PY{n+nx}{MakeClerk}\PY{p}{(}\PY{n+nx}{oldServers}\PY{p}{[}\PY{n+nx}{Rand}\PY{p}{(}\PY{p}{)}\PY{+w}{ }\PY{o}{\PYZpc{}}\PY{+w}{ }\PY{n+nb}{len}\PY{p}{(}\PY{n+nx}{oldServers}\PY{p}{)}\PY{p}{]}\PY{p}{)}
\PY{+w}{  }\PY{n+nx}{oldState}\PY{+w}{ }\PY{o}{:=}\PY{+w}{ }\PY{n+nx}{oldClerk}\PY{p}{.}\PY{n+nx}{GetStateAndSeal}\PY{p}{(}\PY{n+nx}{newEpoch}\PY{p}{)}

\PY{+w}{  }\PY{c+c1}{// make clerks to all of the new servers}
\PY{+w}{  }\PY{k+kd}{var}\PY{+w}{ }\PY{n+nx}{newClerks}\PY{+w}{ }\PY{p}{=}\PY{+w}{ }\PY{n+nb}{make}\PY{p}{(}\PY{p}{[}\PY{p}{]}\PY{n+nx}{Clerk}\PY{p}{,}\PY{+w}{ }\PY{n+nb}{len}\PY{p}{(}\PY{n+nx}{newServers}\PY{p}{)}\PY{p}{)}
\PY{+w}{  }\PY{k}{for}\PY{+w}{ }\PY{n+nx}{i}\PY{+w}{ }\PY{o}{:=}\PY{+w}{ }\PY{l+m+mi}{0}\PY{p}{;}\PY{+w}{ }\PY{n+nx}{i}\PY{+w}{ }\PY{p}{\PYZlt{}}\PY{+w}{ }\PY{n+nb}{len}\PY{p}{(}\PY{n+nx}{newServers}\PY{p}{)}\PY{p}{;}\PY{+w}{ }\PY{n+nx}{i}\PY{o}{++}\PY{+w}{ }\PY{p}{\PYZob{}}
\PY{+w}{    }\PY{n+nx}{newClerks}\PY{p}{[}\PY{n+nx}{i}\PY{p}{]}\PY{+w}{ }\PY{p}{=}\PY{+w}{ }\PY{n+nx}{MakeClerk}\PY{p}{(}\PY{n+nx}{newServers}\PY{p}{[}\PY{n+nx}{i}\PY{p}{]}\PY{p}{)}
\PY{+w}{  }\PY{p}{\PYZcb{}}

\PY{+w}{  }\PY{c+c1}{// set state on all the new servers}
\PY{+w}{  }\PY{n+nx}{wg}\PY{+w}{ }\PY{o}{:=}\PY{+w}{ }\PY{n+nb}{new}\PY{p}{(}\PY{n+nx}{WaitGroup}\PY{p}{)}
\PY{+w}{  }\PY{k}{for}\PY{+w}{ }\PY{n+nx}{i}\PY{+w}{ }\PY{o}{:=}\PY{+w}{ }\PY{l+m+mi}{0}\PY{p}{;}\PY{+w}{ }\PY{n+nx}{i}\PY{+w}{ }\PY{p}{\PYZlt{}}\PY{+w}{ }\PY{n+nb}{len}\PY{p}{(}\PY{n+nx}{newClerks}\PY{p}{)}\PY{p}{;}\PY{+w}{ }\PY{n+nx}{i}\PY{o}{++}\PY{+w}{ }\PY{p}{\PYZob{}}
\PY{+w}{    }\PY{n+nx}{wg}\PY{p}{.}\PY{n+nx}{Add}\PY{p}{(}\PY{l+m+mi}{1}\PY{p}{)}
\PY{+w}{    }\PY{k}{go}\PY{+w}{ }\PY{k+kd}{func}\PY{p}{(}\PY{n+nx}{i}\PY{+w}{ }\PY{k+kt}{int}\PY{p}{)}\PY{+w}{ }\PY{p}{\PYZob{}}
\PY{+w}{      }\PY{n+nx}{newClerks}\PY{p}{[}\PY{n+nx}{i}\PY{p}{]}\PY{p}{.}\PY{n+nx}{SetNewEpochState}\PY{p}{(}\PY{n+nx}{newEpoch}\PY{p}{,}\PY{+w}{ }\PY{n+nx}{oldState}\PY{p}{)}
\PY{+w}{      }\PY{n+nx}{wg}\PY{p}{.}\PY{n+nx}{Done}\PY{p}{(}\PY{p}{)}
\PY{+w}{    }\PY{p}{\PYZcb{}}\PY{p}{(}\PY{n+nx}{i}\PY{p}{)}
\PY{+w}{  }\PY{p}{\PYZcb{}}
\PY{+w}{  }\PY{n+nx}{wg}\PY{p}{.}\PY{n+nx}{Wait}\PY{p}{(}\PY{p}{)}

\PY{+w}{  }\PY{c+c1}{// write new addresses to config service}
\PY{+w}{  }\PY{n+nx}{err}\PY{+w}{ }\PY{o}{:=}\PY{+w}{ }\PY{n+nx}{configClerk}\PY{p}{.}\PY{n+nx}{TryWriteConfig}\PY{p}{(}\PY{n+nx}{newEpoch}\PY{p}{,}\PY{+w}{ }\PY{n+nx}{newServers}\PY{p}{)}
\PY{+w}{  }\PY{k}{if}\PY{+w}{ }\PY{n+nx}{err}\PY{+w}{ }\PY{o}{==}\PY{+w}{ }\PY{k+kc}{nil}\PY{+w}{ }\PY{p}{\PYZob{}}
\PY{+w}{    }\PY{c+c1}{// activate the new primary server}
\PY{+w}{    }\PY{n+nx}{newClerks}\PY{p}{[}\PY{l+m+mi}{0}\PY{p}{]}\PY{p}{.}\PY{n+nx}{BecomePrimary}\PY{p}{(}\PY{n+nx}{newEpoch}\PY{p}{)}
\PY{+w}{  }\PY{p}{\PYZcb{}}
\PY{p}{\PYZcb{}}
\end{Verbatim}

%% file: code/lease.tex
\begin{Verbatim}[commandchars=\\\{\},numbers=left,firstnumber=1,stepnumber=1,codes={\catcode`\$=3\catcode`\^=7\catcode`\_=8\relax},fontsize=\scriptsize,numbersep=6pt,xleftmargin=0.2in]
\PY{k+kd}{func}\PY{+w}{ }\PY{p}{(}\PY{n+nx}{s}\PY{+w}{ }\PY{o}{*}\PY{n+nx}{Server}\PY{p}{)}\PY{+w}{ }\PY{n+nx}{ApplyReadonly}\PY{p}{(}\PY{n+nx}{op}\PY{p}{)}\PY{+w}{ }\PY{n+nx}{Result}\PY{+w}{ }\PY{p}{\PYZob{}}
\PY{+w}{  }\PY{n+nx}{s}\PY{p}{.}\PY{n+nx}{mutex}\PY{p}{.}\PY{n+nx}{Lock}\PY{p}{(}\PY{p}{)}

\PY{+w}{  }\PY{k}{if}\PY{+w}{ }\PY{n+nx}{s}\PY{p}{.}\PY{n+nx}{leaseExpiry}\PY{+w}{ }\PY{p}{\PYZgt{}}\PY{+w}{ }\PY{n+nx}{GetTimeRange}\PY{p}{(}\PY{p}{)}\PY{p}{.}\PY{n+nx}{latest}\PY{+w}{ }\PY{p}{\PYZob{}}
\PY{+w}{    }\PY{n+nx}{e}\PY{+w}{ }\PY{o}{:=}\PY{+w}{ }\PY{n+nx}{s}\PY{p}{.}\PY{n+nx}{epoch}
\PY{+w}{    }\PY{n+nx}{idx}\PY{p}{,}\PY{+w}{ }\PY{n+nx}{res}\PY{+w}{ }\PY{o}{:=}\PY{+w}{ }\PY{n+nx}{s}\PY{p}{.}\PY{n+nx}{stateLogger}\PY{p}{.}\PY{n+nx}{LocalRead}\PY{p}{(}\PY{n+nx}{op}\PY{p}{)}
\PY{+w}{    }\PY{n+nx}{s}\PY{p}{.}\PY{n+nx}{mutex}\PY{p}{.}\PY{n+nx}{Unlock}\PY{p}{(}\PY{p}{)}

\PY{+w}{    }\PY{k}{if}\PY{+w}{ }\PY{n+nx}{s}\PY{p}{.}\PY{n+nx}{waitForCommitted}\PY{p}{(}\PY{n+nx}{e}\PY{p}{,}\PY{+w}{ }\PY{n+nx}{op}\PY{p}{,}\PY{+w}{ }\PY{n+nx}{idx}\PY{p}{)}\PY{+w}{ }\PY{p}{\PYZob{}}
\PY{+w}{      }\PY{k}{return}\PY{+w}{ }\PY{n+nx}{res}
\PY{+w}{    }\PY{p}{\PYZcb{}}\PY{+w}{ }\PY{k}{else}\PY{+w}{ }\PY{p}{\PYZob{}}
\PY{+w}{      }\PY{k}{return}\PY{+w}{ }\PY{n+nx}{ErrRetry}
\PY{+w}{    }\PY{p}{\PYZcb{}}
\PY{+w}{  }\PY{p}{\PYZcb{}}\PY{+w}{ }\PY{k}{else}\PY{+w}{ }\PY{p}{\PYZob{}}
\PY{+w}{    }\PY{n+nx}{s}\PY{p}{.}\PY{n+nx}{mutex}\PY{p}{.}\PY{n+nx}{Unlock}\PY{p}{(}\PY{p}{)}
\PY{+w}{    }\PY{k}{return}\PY{+w}{ }\PY{n+nx}{ErrRetry}
\PY{+w}{  }\PY{p}{\PYZcb{}}
\PY{p}{\PYZcb{}}
\end{Verbatim}

%% file: code/paxos.tex
\begin{Verbatim}[commandchars=\\\{\},numbers=left,firstnumber=1,stepnumber=1,codes={\catcode`\$=3\catcode`\^=7\catcode`\_=8\relax},fontsize=\scriptsize,numbersep=6pt,xleftmargin=0.2in]
\PY{k+kd}{func}\PY{+w}{ }\PY{p}{(}\PY{n+nx}{p}\PY{+w}{ }\PY{o}{*}\PY{n+nx}{Paxos}\PY{p}{)}\PY{+w}{ }\PY{n+nx}{WeakRead}\PY{p}{(}\PY{p}{)}\PY{+w}{ }\PY{p}{[}\PY{p}{]}\PY{k+kt}{byte}
\PY{k+kd}{func}\PY{+w}{ }\PY{p}{(}\PY{n+nx}{p}\PY{+w}{ }\PY{o}{*}\PY{n+nx}{Paxos}\PY{p}{)}\PY{+w}{ }\PY{n+nx}{Begin}\PY{p}{(}\PY{p}{)}\PY{+w}{ }\PY{p}{(}\PY{n+nx}{oldstate}\PY{+w}{ }\PY{p}{[}\PY{p}{]}\PY{k+kt}{byte}\PY{p}{,}
\PY{+w}{                         }\PY{n+nx}{commit}\PY{+w}{ }\PY{k+kd}{func}\PY{p}{(}\PY{n+nx}{newstate}\PY{+w}{ }\PY{p}{[}\PY{p}{]}\PY{k+kt}{byte}\PY{p}{)}\PY{+w}{ }\PY{k+kt}{error}\PY{p}{)}
\PY{k+kd}{func}\PY{+w}{ }\PY{p}{(}\PY{n+nx}{p}\PY{+w}{ }\PY{o}{*}\PY{n+nx}{Paxos}\PY{p}{)}\PY{+w}{ }\PY{n+nx}{TryBecomingLeader}\PY{p}{(}\PY{p}{)}
\end{Verbatim}

%% file: code/configimpl.tex
\begin{Verbatim}[commandchars=\\\{\},numbers=left,firstnumber=1,stepnumber=1,codes={\catcode`\$=3\catcode`\^=7\catcode`\_=8\relax},fontsize=\scriptsize,numbersep=6pt,xleftmargin=0.2in]
\PY{k+kd}{func}\PY{+w}{ }\PY{p}{(}\PY{n+nx}{c}\PY{+w}{ }\PY{o}{*}\PY{n+nx}{ConfigService}\PY{p}{)}\PY{+w}{ }\PY{n+nx}{ReserveEpochAndGetConfig}\PY{p}{(}\PY{n+nx}{args}\PY{+w}{ }\PY{p}{[}\PY{p}{]}\PY{k+kt}{byte}\PY{p}{,}
\PY{+w}{                                                 }\PY{n+nx}{reply}\PY{+w}{ }\PY{o}{*}\PY{p}{[}\PY{p}{]}\PY{k+kt}{byte}\PY{p}{)}\PY{+w}{ }\PY{p}{\PYZob{}}
\PY{+w}{  }\PY{n+nx}{oldstate}\PY{p}{,}\PY{+w}{ }\PY{n+nx}{commit}\PY{+w}{ }\PY{o}{:=}\PY{+w}{ }\PY{n+nx}{c}\PY{p}{.}\PY{n+nx}{paxos}\PY{p}{.}\PY{n+nx}{Begin}\PY{p}{(}\PY{p}{)}
\PY{+w}{  }\PY{n+nx}{st}\PY{+w}{ }\PY{o}{:=}\PY{+w}{ }\PY{n+nx}{unmarshal}\PY{p}{(}\PY{n+nx}{oldstate}\PY{p}{)}
\PY{+w}{  }\PY{n+nx}{st}\PY{p}{.}\PY{n+nx}{reservedEpoch}\PY{+w}{ }\PY{p}{=}\PY{+w}{ }\PY{n+nx}{st}\PY{p}{.}\PY{n+nx}{reservedEpoch}\PY{+w}{ }\PY{o}{+}\PY{+w}{ }\PY{l+m+mi}{1}
\PY{+w}{  }\PY{n+nx}{newstate}\PY{+w}{ }\PY{o}{:=}\PY{+w}{ }\PY{n+nx}{marshal}\PY{p}{(}\PY{n+nx}{st}\PY{p}{)}

\PY{+w}{  }\PY{n+nx}{err}\PY{+w}{ }\PY{o}{:=}\PY{+w}{ }\PY{n+nx}{commit}\PY{p}{(}\PY{n+nx}{newstate}\PY{p}{)}
\PY{+w}{  }\PY{k}{if}\PY{+w}{ }\PY{n+nx}{err}\PY{+w}{ }\PY{o}{!=}\PY{+w}{ }\PY{k+kc}{nil}\PY{+w}{ }\PY{p}{\PYZob{}}
\PY{+w}{    }\PY{o}{*}\PY{n+nx}{reply}\PY{+w}{ }\PY{p}{=}\PY{+w}{ }\PY{n+nx}{marshal}\PY{p}{.}\PY{n+nx}{WriteInt}\PY{p}{(}\PY{k+kc}{nil}\PY{p}{,}\PY{+w}{ }\PY{n+nx}{STAT\PYZus{}ERROR}\PY{p}{)}
\PY{+w}{  }\PY{p}{\PYZcb{}}\PY{+w}{ }\PY{k}{else}\PY{+w}{ }\PY{p}{\PYZob{}}
\PY{+w}{    }\PY{o}{*}\PY{n+nx}{reply}\PY{+w}{ }\PY{p}{=}\PY{+w}{ }\PY{n+nx}{marshal}\PY{p}{.}\PY{n+nx}{WriteInt}\PY{p}{(}\PY{k+kc}{nil}\PY{p}{,}\PY{+w}{ }\PY{n+nx}{STAT\PYZus{}OK}\PY{p}{)}
\PY{+w}{    }\PY{o}{*}\PY{n+nx}{reply}\PY{+w}{ }\PY{p}{=}\PY{+w}{ }\PY{n+nx}{marshal}\PY{p}{.}\PY{n+nx}{WriteInt}\PY{p}{(}\PY{o}{*}\PY{n+nx}{reply}\PY{p}{,}\PY{+w}{ }\PY{n+nx}{st}\PY{p}{.}\PY{n+nx}{reservedEpoch}\PY{p}{)}
\PY{+w}{    }\PY{o}{*}\PY{n+nx}{reply}\PY{+w}{ }\PY{p}{=}\PY{+w}{ }\PY{n+nx}{marshal}\PY{p}{.}\PY{n+nx}{WriteBytes}\PY{p}{(}\PY{o}{*}\PY{n+nx}{reply}\PY{p}{,}\PY{+w}{ }\PY{n+nx}{encode\PYZus{}cfg}\PY{p}{(}\PY{n+nx}{st}\PY{p}{.}\PY{n+nx}{config}\PY{p}{)}\PY{p}{)}
\PY{+w}{  }\PY{p}{\PYZcb{}}
\PY{p}{\PYZcb{}}
\end{Verbatim}

%% file: code/vsm.tex
\begin{Verbatim}[commandchars=\\\{\},numbers=left,firstnumber=1,stepnumber=1,codes={\catcode`\$=3\catcode`\^=7\catcode`\_=8\relax},fontsize=\scriptsize,numbersep=6pt,xleftmargin=0.2in]
\PY{k+kd}{type}\PY{+w}{ }\PY{n+nx}{VersionedStateMachine}\PY{+w}{ }\PY{k+kd}{struct}\PY{+w}{ }\PY{p}{\PYZob{}}
\PY{+w}{  }\PY{n+nx}{Apply}\PY{+w}{    }\PY{k+kd}{func}\PY{p}{(}\PY{n+nx}{op}\PY{+w}{ }\PY{p}{[}\PY{p}{]}\PY{k+kt}{byte}\PY{p}{,}\PY{+w}{ }\PY{n+nx}{idx}\PY{+w}{ }\PY{k+kt}{uint64}\PY{p}{)}\PY{+w}{ }\PY{p}{[}\PY{p}{]}\PY{k+kt}{byte}
\PY{+w}{  }\PY{n+nx}{Read}\PY{+w}{     }\PY{k+kd}{func}\PY{p}{(}\PY{n+nx}{op}\PY{+w}{ }\PY{p}{[}\PY{p}{]}\PY{k+kt}{byte}\PY{p}{)}\PY{+w}{ }\PY{p}{(}\PY{k+kt}{uint64}\PY{p}{,}\PY{+w}{ }\PY{p}{[}\PY{p}{]}\PY{k+kt}{byte}\PY{p}{)}
\PY{+w}{  }\PY{n+nx}{SetState}\PY{+w}{ }\PY{k+kd}{func}\PY{p}{(}\PY{n+nx}{snap}\PY{+w}{ }\PY{p}{[}\PY{p}{]}\PY{k+kt}{byte}\PY{p}{,}\PY{+w}{ }\PY{n+nx}{idx}\PY{+w}{ }\PY{k+kt}{uint64}\PY{p}{)}
\PY{+w}{  }\PY{n+nx}{GetState}\PY{+w}{ }\PY{k+kd}{func}\PY{p}{(}\PY{p}{)}\PY{+w}{ }\PY{p}{[}\PY{p}{]}\PY{k+kt}{byte}
\PY{p}{\PYZcb{}}
\end{Verbatim}

%% file: code/rsmclerk.tex
\begin{Verbatim}[commandchars=\\\{\},numbers=left,firstnumber=1,stepnumber=1,codes={\catcode`\$=3\catcode`\^=7\catcode`\_=8\relax},fontsize=\scriptsize,numbersep=6pt,xleftmargin=0.2in]
\PY{k+kd}{func}\PY{+w}{ }\PY{n+nx}{MakeClerk}\PY{p}{(}\PY{n+nx}{confAddr}\PY{+w}{ }\PY{n+nx}{Address}\PY{p}{)}\PY{+w}{ }\PY{o}{*}\PY{n+nx}{Clerk}
\PY{k+kd}{func}\PY{+w}{ }\PY{p}{(}\PY{n+nx}{ck}\PY{+w}{ }\PY{o}{*}\PY{n+nx}{Clerk}\PY{p}{)}\PY{+w}{ }\PY{n+nx}{Apply}\PY{p}{(}\PY{n+nx}{op}\PY{+w}{ }\PY{p}{[}\PY{p}{]}\PY{k+kt}{byte}\PY{p}{)}\PY{+w}{ }\PY{p}{[}\PY{p}{]}\PY{k+kt}{byte}
\PY{k+kd}{func}\PY{+w}{ }\PY{p}{(}\PY{n+nx}{ck}\PY{+w}{ }\PY{o}{*}\PY{n+nx}{Clerk}\PY{p}{)}\PY{+w}{ }\PY{n+nx}{Read}\PY{p}{(}\PY{n+nx}{op}\PY{+w}{ }\PY{p}{[}\PY{p}{]}\PY{k+kt}{byte}\PY{p}{)}\PY{+w}{ }\PY{p}{[}\PY{p}{]}\PY{k+kt}{byte}
\end{Verbatim}

%% file: code/kvclerk.tex
\begin{Verbatim}[commandchars=\\\{\},numbers=left,firstnumber=1,stepnumber=1,codes={\catcode`\$=3\catcode`\^=7\catcode`\_=8\relax},fontsize=\scriptsize,numbersep=6pt,xleftmargin=0.2in]
\PY{k+kd}{func}\PY{+w}{ }\PY{p}{(}\PY{n+nx}{ck}\PY{+w}{ }\PY{o}{*}\PY{n+nx}{Clerk}\PY{p}{)}\PY{+w}{ }\PY{n+nx}{Put}\PY{p}{(}\PY{n+nx}{key}\PY{p}{,}\PY{+w}{ }\PY{n+nx}{val}\PY{+w}{ }\PY{k+kt}{string}\PY{p}{)}
\PY{k+kd}{func}\PY{+w}{ }\PY{p}{(}\PY{n+nx}{ck}\PY{+w}{ }\PY{o}{*}\PY{n+nx}{Clerk}\PY{p}{)}\PY{+w}{ }\PY{n+nx}{CondPut}\PY{p}{(}\PY{n+nx}{key}\PY{p}{,}\PY{+w}{ }\PY{n+nx}{expect}\PY{p}{,}\PY{+w}{ }\PY{n+nx}{val}\PY{+w}{ }\PY{k+kt}{string}\PY{p}{)}
\PY{k+kd}{func}\PY{+w}{ }\PY{p}{(}\PY{n+nx}{ck}\PY{+w}{ }\PY{o}{*}\PY{n+nx}{Clerk}\PY{p}{)}\PY{+w}{ }\PY{n+nx}{Get}\PY{p}{(}\PY{n+nx}{key}\PY{+w}{ }\PY{k+kt}{string}\PY{p}{)}\PY{+w}{ }\PY{k+kt}{string}
\end{Verbatim}

%% file: code/leasekv.tex
\begin{Verbatim}[commandchars=\\\{\},numbers=left,firstnumber=1,stepnumber=1,codes={\catcode`\$=3\catcode`\^=7\catcode`\_=8\relax},fontsize=\scriptsize,numbersep=6pt,xleftmargin=0.2in]
\PY{k+kd}{func}\PY{+w}{ }\PY{p}{(}\PY{n+nx}{k}\PY{+w}{ }\PY{o}{*}\PY{n+nx}{CacheKv}\PY{p}{)}\PY{+w}{ }\PY{n+nx}{GetAndCache}\PY{p}{(}\PY{n+nx}{key}\PY{+w}{ }\PY{k+kt}{string}\PY{p}{,}
\PY{+w}{                              }\PY{n+nx}{cachetime}\PY{+w}{ }\PY{k+kt}{uint64}\PY{p}{)}\PY{+w}{ }\PY{k+kt}{string}\PY{+w}{ }\PY{p}{\PYZob{}}
\PY{+w}{  }\PY{k}{for}\PY{+w}{ }\PY{p}{\PYZob{}}
\PY{+w}{    }\PY{n+nx}{old}\PY{+w}{ }\PY{o}{:=}\PY{+w}{ }\PY{n+nx}{k}\PY{p}{.}\PY{n+nx}{kv}\PY{p}{.}\PY{n+nx}{Get}\PY{p}{(}\PY{n+nx}{key}\PY{p}{)}
\PY{+w}{    }\PY{n+nx}{new}\PY{+w}{ }\PY{o}{:=}\PY{+w}{ }\PY{n+nx}{old}

\PY{+w}{    }\PY{n+nx}{newExpiration}\PY{+w}{ }\PY{o}{:=}\PY{+w}{ }\PY{n+nb}{max}\PY{p}{(}\PY{n+nx}{GetTimeRange}\PY{p}{(}\PY{p}{)}\PY{p}{.}\PY{n+nx}{latest}\PY{o}{+}\PY{n+nx}{cachetime}\PY{p}{,}
\PY{+w}{      }\PY{n+nx}{old}\PY{p}{.}\PY{n+nx}{leaseExpiration}\PY{p}{)}
\PY{+w}{    }\PY{n+nx}{new}\PY{p}{.}\PY{n+nx}{leaseExpiration}\PY{+w}{ }\PY{p}{=}\PY{+w}{ }\PY{n+nx}{newExpiration}

\PY{+w}{    }\PY{c+c1}{// Try to update the lease expiration time on the backend}
\PY{+w}{    }\PY{n+nx}{resp}\PY{+w}{ }\PY{o}{:=}\PY{+w}{ }\PY{n+nx}{k}\PY{p}{.}\PY{n+nx}{kv}\PY{p}{.}\PY{n+nx}{CondPut}\PY{p}{(}\PY{n+nx}{key}\PY{p}{,}\PY{+w}{ }\PY{n+nx}{old}\PY{p}{,}\PY{+w}{ }\PY{n+nx}{new}\PY{p}{)}
\PY{+w}{    }\PY{k}{if}\PY{+w}{ }\PY{n+nx}{resp}\PY{+w}{ }\PY{o}{==}\PY{+w}{ }\PY{l+s}{\PYZdq{}ok\PYZdq{}}\PY{+w}{ }\PY{p}{\PYZob{}}
\PY{+w}{      }\PY{n+nx}{k}\PY{p}{.}\PY{n+nx}{mu}\PY{p}{.}\PY{n+nx}{Lock}\PY{p}{(}\PY{p}{)}
\PY{+w}{      }\PY{n+nx}{k}\PY{p}{.}\PY{n+nx}{cache}\PY{p}{[}\PY{n+nx}{key}\PY{p}{]}\PY{+w}{ }\PY{p}{=}\PY{+w}{ }\PY{n+nx}{cacheValue}\PY{p}{\PYZob{}}\PY{n+nx}{v}\PY{p}{:}\PY{+w}{ }\PY{n+nx}{old}\PY{p}{.}\PY{n+nx}{v}\PY{p}{,}\PY{+w}{ }\PY{n+nx}{l}\PY{p}{:}\PY{+w}{ }\PY{n+nx}{newLeaseExpiration}\PY{p}{\PYZcb{}}
\PY{+w}{      }\PY{n+nx}{k}\PY{p}{.}\PY{n+nx}{mu}\PY{p}{.}\PY{n+nx}{Unlock}\PY{p}{(}\PY{p}{)}
\PY{+w}{      }\PY{k}{return}\PY{+w}{ }\PY{n+nx}{old}\PY{p}{.}\PY{n+nx}{v}
\PY{+w}{    }\PY{p}{\PYZcb{}}
\PY{+w}{  }\PY{p}{\PYZcb{}}
\PY{p}{\PYZcb{}}
\end{Verbatim}

%% file: code/lockservice.tex
\begin{Verbatim}[commandchars=\\\{\},numbers=left,firstnumber=1,stepnumber=1,codes={\catcode`\$=3\catcode`\^=7\catcode`\_=8\relax},fontsize=\scriptsize,numbersep=6pt,xleftmargin=0.2in]
\PY{k+kd}{func}\PY{+w}{ }\PY{p}{(}\PY{n+nx}{ck}\PY{+w}{ }\PY{o}{*}\PY{n+nx}{Clerk}\PY{p}{)}\PY{+w}{ }\PY{n+nx}{Acquire}\PY{p}{(}\PY{n+nx}{lk}\PY{+w}{ }\PY{k+kt}{string}\PY{p}{)}\PY{+w}{ }\PY{o}{*}\PY{n+nx}{Locked}
\PY{k+kd}{func}\PY{+w}{ }\PY{p}{(}\PY{n+nx}{l}\PY{+w}{ }\PY{o}{*}\PY{n+nx}{Locked}\PY{p}{)}\PY{+w}{ }\PY{n+nx}{Release}\PY{p}{(}\PY{p}{)}
\end{Verbatim}

%% file: design.tex
\section{\sys}
\label{sec:design}

To formally verify distributed systems such as the case studies
described in the previous section, \sys{} adopts the ideas of concurrent
separation logic (CSL)~\cite{ohearn:csl, jung:iris-jfp}.  CSL enables
modular specifications and proofs: a developer can take two verified
components, each with their own specification, and use both of them in
their application without worrying that the combination breaks either
component's proof.  In the context of distributed systems, this allows
\sys{} developers to separately specify and verify different services
that run on different machines but that will eventually be used together
(e.g., a configuration service, a key-value store, and a lock service),
as well as different libraries that will run on the same machine (e.g., a
clerk that talks to \kv{}, a clerk that talks to the lock service, etc).

In the rest of this section, we first introduce \sys{}'s
execution model (\autoref{sec:design:model}), followed by how \sys{}
generalizes separation logic and resource ownership to distributed systems
(\autoref{sec:design:csl}), and \sys{}'s library of reasoning principles.
\showto{conf}{The extended version of the paper
provides additional details, such as \sys{}'s handling of
exactly-once operations~\xref{sec:design:exactlyonce} and
crashes~\xref{sec:design:chl}.}

\subsection{Execution model}
\label{sec:design:model}

\sys{} models distributed systems as a collection of nodes, each running
a multithreaded program written in Go.  Each node has its own memory
heap (accessed in Go using loads and stores) as well as durable storage
(accessed by reading from and writing to files using \cc{read()} and
\cc{write()}).

\paragraph{Crash recovery.}
Each node has a \cc{main()} function that runs when the
node starts up for the first time as well as when the node restarts after
a crash.  When a node crashes, it loses the contents of its memory heap
and restarts with an empty heap, but retains its durable state.

Nodes crash independently of one another. A few nodes might crash while
others keep running, or all of the nodes might crash at the same time.
Crashes can happen at any point, including when a node is still recovering
from an earlier crash.  For instance, a node's \cc{main()} function might
have a recovery phase during which it loads durable state into memory or
communicates with other nodes to restore its state; crashes can occur
even during this phase.

\paragraph{Unreliable network.}
Nodes communicate over an unreliable network. The low-level network API
has a notion of a \cc{Connection}, resembling a
connected UDP socket. The API provides two functions, \cc{conn.Send(msg)}
and \cc{conn.Receive()} that respectively send and receive messages
over that connection. \sys{} models the network as unreliable:
\cc{conn.Send(msg)} is not guaranteed to deliver messages in order,
and messages may be dropped or duplicated.

\paragraph{Clocks.}
There is a global clock, which advances monotonically and represents
a notion of wall-clock time.  Every node exposes a
TrueTime-like API~\cite{corbett:spanner}, \cc{GetTimeRange()}, which
returns a pair of timestamps that represent an interval (lower and upper
bounds) that, according to our model, must contain the global clock value.
This assumes that node clocks are synchronized to within known bounds (on
the order of less than a second, for the purposes of \kv{}'s use of leases).

\subsection{Separation logic for distributed systems}
\label{sec:design:csl}

\sys{} generalizes concurrent separation logic (CSL)~\cite{ohearn:csl, jung:iris-jfp}
to reason about distributed
systems. CSL uses Hoare logic-style specifications for pieces of code
(e.g., functions) of the form $\hoare{P}{\cc{f()}}{Q}$ meaning
the precondition for running $\cc{f()}$ is the assertion $P$ and the postcondition is $Q$.
To prove such a spec, a developer applies proof rules
to reason line-by-line about \cc{f()},
starting with a state matching $P$, and showing that the final state matches $Q$.

This section reviews the background on CSL and introduces key abstractions
that \sys{} provides on top of CSL, along with how they are used in
\kv{}'s proof.

\subsubsection{Ownership reasoning}

In separation logic, assertions not only describe what is true
about a system's state, but also what parts of the state are
logically \emph{owned} by the thread executing the code at that point.
For example, the assertion $x \mapsto v$ (pronounced ``$x$ points to
$v$'') says that memory location $x$ stores a value $v$ \emph{and} that
the thread running that function owns the location $x$, in the sense that,
as long as this thread continues to own this assertion, no other thread
can access location $x$.  Such ownership constraints form the basis for
modular reasoning: for instance, the fact that no other thread can access
location $x$ allows a developer to reason about this function without
considering other concurrently executing code.

\sys{}'s library brings this ownership-based modular reasoning to
distributed systems.  \sys{} provides per-node heap points-to resources:
$x \mapsto_j v$ denotes ownership of location $x$ with value $v$ on node
$j$'s heap.  \sys{} also provides resources for network state and file
contents, as we discuss later.  In a distributed systems setting,
this enables the developer to verify the code running on one node without
worrying about what code might be running on other nodes at the same time.

Separation logic additionally introduces a new logical connective, $\ast$,
called \emph{separating conjunction}.  The assertion $P \ast Q$ holds
in a state $s$ if both $P$ and $Q$ are true in $s$, and furthermore, $s$
can be split into two \emph{disjoint} resources satisfying $P$ and $Q$,
respectively. In conventional CSL, disjointedness means separate subsets
of a program's memory heap.  \sys{} uses the separation conjunction to
account for separation across different nodes as well.

\subsubsection{Ghost resources}
\label{sec:design:ghost}

In addition to physical resources like the heap, separation logic allows
proofs to use \emph{ghost resources}, a modern form of auxiliary variables~\cite{Kleymann:99,Jones2010TheRO}.
Ghost resources talk about the state of
the system at a more abstract level.
In concurrent separation logic, ghost resources represent ghost
state---state that is not materialized by the actual running code, but
is useful for specification and proof.  Just like physical resources,
ghost resources can be owned.  While the evolution of physical resources
is entirely determined by the code (e.g., based on how the code modifies
memory or file contents), ghost resources are controlled by the proof.

Ghost resources are especially useful for reasoning about distributed
systems because they can span nodes and allow developers to reason about
the system at a higher level of abstraction.  Ghost resources are more
powerful than regular abstract state, because these resources can be owned,
which in turn provides constraints on how different threads can modify
the ghost state, and thereby enables modular reasoning.

\paragraph{Epochs.}
Using ghost resources, \sys{} provides an epoch abstraction, which is used in
the proof of \kv{} to keep track of and reason about the current configuration.
\sys{} provides two resources for representing epochs.  The first,
$\currentepoch{e}$, states that the current epoch number is exactly $e$.  This
resource is owned by the configuration service: it is the only component that
can approve a reconfiguration.  The second, $\currentepochlb{e}$, states that
the current epoch is at least $e$.  This resource is \emph{duplicable}, meaning
that many threads can have it at the same time.  In a way, this resource
represents \emph{knowledge} of the fact that the epoch is at least $e$, rather
than any exclusive ownership of some part of the state.  The fact that
$\currentepochlb{e}$ is duplicable implies that epoch numbers are monotonically
increasing (i.e., the resource promises that the current epoch number cannot
decrease).  Many \kv{} components, including the primary and the backup
replicas, make use of this resource to represent knowledge that a new epoch
exists.  This means that a server can reject operations from earlier epochs,
such as a stale \cc{SetNewEpochState()}.

\paragraph{Logs.}
\sys{} provides a log abstraction using ghost resources,
encoded as an append-only list.  The proof of \kv{} encodes the main
logical state of each replica server using this log abstraction,
representing the operations that the node has applied so far.  There are
three kinds of ghost resources provided by \sys{} that talk about the
state of an append-only log:

The \emph{points-to} resource $\aol{a}{\ell}$ denotes ownership of an
append-only list named $a$ with current value $\ell$. The only way to
update this points-to resource in the proof is to go from ownership of
$\aol{a}{\ell}$ to $\aol{a}{\ell + \ell'}$, i.e. to append at the end.

The \emph{lower bound} resource $\aolb{a}{\ell}$ denotes knowledge that
the list $a$ has prefix $\ell$.
This is similar to the lower-bound epoch resource $\currentepochlb{e}$ described above,
and just like it, $\aolb{a}{\ell}$ is duplicable.
Other parts of the proof cannot possibly violate lower bound resources.

Finally, the \emph{read-only} resource $\aolro{a}{\ell}$ denotes knowledge
that $a$ has value $\ell$ and can never be updated.  To establish
this read-only resource, a proof has to give up ownership of
$\aol{a}{\ell}$ and in exchange get ownership of the read-only
resource. After this, no part of the proof can possibly have ownership
of $\aol{a}{\ell}$ so the list can never be updated again.

\kv{}'s proof represents operations accepted by each server as of epoch
number $e$ with append-only lists: server $j \in \{0, 1, \ldots, n\}$
owns the resource $\aol{\cc{accepted}_j[e]}{\ell}$.\footnote{
The labels are such that server 0 is the primary and the rest are backups.
}
Server $j$ can only gain knowledge (not ownership) about other servers'
$\cc{accepted}_k[e]$ list and does this through RPCs (discussed in
\autoref{sec:design:rpc}).
All servers also own heap resources for their in-memory representation of this
abstract state, but the proof does not involve sharing these heap resources
across nodes. Servers only talk about other servers in terms of ghost resources
for their $\cc{accepted}_j[e]$ list. Read-only resources are used to represent
sealed replicas.

The proof also has a global points-to $\aol{\cc{committed}}{\ell}$ that
represents the committed list of operations.  When a primary server
commits an operation, the proof updates the \cc{committed} points-to
resource. However, when reconfiguration happens, the new primary server
will need to do the same. Thus, the \cc{committed} points-to resource
cannot be permanently owned by any one node. To share this resource between
nodes, the proof uses a separation logic invariant,
which we explain next.

\subsubsection{Invariants}
\label{sec:design:invariants}

Concurrent separation logic allows for resources to be shared through \emph{invariants}, which can
talk about the resources relevant to only a small part of the system without
needing to know about the entire system's state.%
\footnote{Different variants of CSL come with different flavors of invariants.
Here, we are explaining invariants as they work in Iris~\cite{jung:iris-jfp}.}
These invariants maintain ownership of resources that
must always be available. The assertion $\knowInv{}{P}$ denotes an invariant
that maintains ownership of $P$.
When reasoning about code, proofs can temporarily ``open'' $\knowInv{}{P}$ to get ownership
of $P$, but are required within one physically atomic step of the code
to return ownership of $P$ in order to ``close'' the invariant. An
invariant is created by starting with ownership of $P$ and giving it up to
establish $\knowInv{}{P}$.  The proposition $\knowInv{}{P}$ asserts
\emph{knowledge} of the invariant, as opposed to direct ownership of
the resources $P$; many threads can hold $\knowInv{}{P}$ at the same time.

Multiple invariants that each talk about separate parts of a larger
system can be freely combined. As an example, a per-node invariant can
describe how resources are shared between the node's threads; this is
how invariants are used in traditional single-machine CSL
(\cite{ohearn:csl,jung:iris-jfp}).  At the same time, a separate invariant can connect the
logical state of all replicas to ensure that the replicas agree on the
log of accepted operations.  Finally, yet another invariant can cover the
configuration service and how the reconfiguration logic ensures that
only one set of servers is active at a time.

\paragraph{Example.}
In the \kv{} proof, each node has a local invariant $I_{\textit{node}_j}$
that maintains ownership of local heap resources and
$\aol{\cc{accepted}_j[e]}{\ell}$ to help reason about node-local concurrency,
using \sys{}'s log ghost resource.

Separately, to reason about how nodes coordinate with each other, the proof has
a ``replication invariant'' $I_{\textit{rep}}$ defined as
$$
\boxed{
\begin{array}{@{}r@{~}l@{}}
\exists \ell, \exists e, &
  \aol {\cc{committed}} {\ell} \ast \\
& \left(\aolb{\cc{accepted}_0[e]} {\ell}\right) \ast
  \cdots \ast
  \left(\aolb{\cc{accepted}_n[e]} {\ell}\right)
\end{array}
}
$$

Here, $\ast$ is the ``separating conjunction'' operator that combines ownership of multiple disjoint
resources.  The invariant maintains ownership of the committed list
of operations and, for every replica server, knowledge that the server
has accepted all the committed operations. This invariant encodes the
primary/backup protocol: in order for an operation to be committed,
all the servers must have accepted it.  Not shown is a part of this invariant
that says epoch $e$ corresponds to a configuration consisting of servers
$0$ through $n$.

When the primary commits an operation $\cc{op}$, the proof opens the replication
invariant to update the \cc{committed} points-to from $\ell$ to $\ell +
[\cc{op}]$. In order to close the invariant after the update, the primary
needs knowledge of the lower-bound resources $\aolb{\cc{accepted}_j[e]}{\ell +
  [op]}$ from all servers $j$. To get these lower bound resources from the
backups to the primary, the proof uses \sys{}'s RPC reasoning principles.

\subsection{Reasoning about RPCs}
\label{sec:design:rpc}

Building on \sys{}'s network model, the \cc{rpc} verified Go RPC
library provides reasoning principles
that allow developers to reason about RPCs much like how they would
reason about local function calls in separation logic.  Key to this RPC
specification is its use of duplicable assertions.  Formally,
an assertion $P$ is called duplicable if $P$ implies $P \ast P$,
meaning that it's possible to create a copy of any resource in $P$.
For instance, $\aol{a}{\ell}$ is not duplicable because one thread's
ownership of this resource precludes any other thread from owning it.
On the other hand, knowledge such as $\aolb{a}{\ell}$ is duplicable.

The notion of duplicability allows stating \sys{}'s RPC specification: for any
function \cc{f} with specification $\hoare{P}{\cc{f}}{Q}$, the specification for
invoking \cc{f} through an RPC is $\hoare{P}{\cc{rpcClient.Call("f")}}{Q}$, as
long as $P$ is duplicable.  Duplicability of $P$ is crucial because the RPC
library may retransmit its request multiple times before it receives a response
and each execution of \cc{f} will consume one instance of $P$.
Note that the specification does not, strictly speaking, require $Q$ to be
duplicable, because \sys{} obtains a fresh copy of $Q$ from each invocation of
\cc{f()} on the server.

\paragraph{Example.}
As an example, consider the \cc{ApplyAsBackup} RPC in \kv{}, issued
by the primary to backup servers when replicating a
new operation.
The postcondition of \cc{ApplyAsBackupRPC(e, index, op)} to server $j$
is the assertion ${\aolb{\cc{accepted}_j[e]}{\ell + [\cc{op}]}}$.
This represents a promise that server $j$ accepted $\cc{op}$ in its log.
As the primary collects more of these lower-bound resources, it will eventually
have enough to commit the operation using the replication invariant $I_{\textit{rep}}$.

When it comes to choosing the precondition of \cc{ApplyAsBackupRPC(e, index, op)}, one might naively pick
``ownership of resources to apply operation \cc{op} once.''  But, such a precondition is
not duplicable, as required by RPCs.
A recurring pattern when specifying RPCs with \sys{} is rephrasing such
preconditions to not involve any exclusive ownership of resources, but instead
talk about knowledge.
The (correct) precondition for the \cc{ApplyAsBackup} RPC is ``knowledge that
\cc{op} is the operation at position \cc{index}.'' The full spec (ignoring
errors) for an \cc{ApplyAsBackupRPC} is:
$$
\hoareV{\text{knowledge that \cc{op} is operation at \cc{index}}}
  {\cc{server}_j.\cc{ApplyAsBackupRPC(e, index, op)}}{\aolb{\cc{accepted}_j[e]}{\ell + [\cc{op}]}}
$$

This precondition allows the primary to retry RPCs to
ensure that every backup has learned about the operation.

\subsection{Reasoning about leases}
\label{sec:design:leases}

To reason about leases, the \sys{} library provides the notion of a
\emph{time-bounded invariant}.  The invariant contains some resources
$R$ representing what the lease $L$ promises to maintain until
its expiration, denoted by $\lease{L}{R}$, and a separate resource
representing the expiration time $\textit{exp}$ of the lease, denoted
by $\leaseexpiration{L}{\textit{exp}}$.  $L$ is a logical identifier
for the lease, and does not show up in execution.

As an example,
\kv{} uses a lease to promise that the configuration epoch number will remain the same,
which ensures that no
reconfiguration will take place for the duration of the lease (which in
turn allows replicas to handle read-only requests on their own).  When the
configuration service hands out such a lease, it creates a time-bounded
invariant $\lease{L}{\currentepoch{e}}$, along with a resource
indicating when the lease expires, $\leaseexpiration{L}{\textit{exp}}$.
It then gives out the invariant and a duplicable version of the lease
expiration resource, $\leaseexpirationlb{L}{\textit{exp}}$; having a
duplicable lower bound on the expiration time, as opposed to the exact
expiration time, simplifies lease renewal.

There are four rules for time-bounded invariants.  First, a time-bounded
invariant can be \emph{created} by giving up ownership of some resource
$R$, and specifying a time at which it will expire.  For instance,
\kv{}'s configuration service does this when issuing a lease in response
to a \cc{GetLease()} RPC, giving up its ownership of $\currentepoch{e}$.

Next, if a time-bounded invariant \emph{expires}, according to the
$\leaseexpiration{L}{\textit{exp}}$ resource, its resources can
be reclaimed.  \kv{}'s configuration service does this as part of
reconfiguration: \cc{TryWriteConfig()} waits for lease expiration, and
gets back ownership of $\currentepoch{e}$, which it can then increment
to $\currentepoch{e+1}$.

Third, a time-bounded invariant can be \emph{extended}: in \kv{}, the
configuration service owns $\leaseexpiration{L}{\textit{exp}}$ if there
is an existing lease, and if another \cc{GetLease()} RPC arrives, the
configuration service extends the lease by advancing the expiration time
to $\leaseexpiration{L}{\textit{exp}+\Delta}$ (and sends a duplicable
$\leaseexpirationlb{L}{\textit{exp}+\Delta}$ to the caller).

Finally, the resources inside of the time-bounded invariant can be
accessed by \emph{opening} the invariant, as long as the time-bounded
invariant has not expired.  Opening a time-bounded invariant comes
with the same obligations as opening a regular invariant---that is,
the proof gets ownership of the resources from the invariant, but must
return them back to close the invariant after at most one atomic step.
In \kv{}, the $\currentepoch{e}$ resource inside the lease invariant is
accessed by the primary-backup replication library (in contrast with the
previous three operations, which all happen on the configuration service).

\subsection{Reasoning about clocks}
\label{sec:design:clocks}

Consider the lease expiration check in \kv{} shown in
\autoref{fig:lease_code}.  This pattern is tricky to reason about: by the
time \cc{s.stateLogger.LocalApplyReadonly()} runs, the lease may no longer
be valid, if there was a long delay right after \cc{if} statement's check.
As a result, whatever invariant the lease was protecting might no longer
be true by the time the developer wants to use it in their proof.

To address this proof challenge, \sys{}'s specification for
\cc{GetTimeRange()} allows the developer to perform arbitrary proof steps
(such as opening and closing invariants and updating ghost resources)
at the instant when \cc{GetTimeRange()} executes.
In the context of these proof steps, the developer also gets access to
a $\currenttime{t}$ resource which represents the current time, and a
promise that the return value $r$ of \cc{GetTimeRange()} satisfies
$r.\mathtt{earliest} \le t \le r.\mathtt{latest}$.  (\sys{}
implements this using logical atomicity~\cite{jacobs:logatom}.)

One subtlety is that, at the instant that \cc{GetTimeRange()} executes,
the code has not yet executed the comparison checking if the lease is
still valid (i.e., comparing to \cc{leaseExpiry}).  As a result, once the
proof gets the $\currenttime{t}$ resource, the developer must explicitly
consider two cases at the instant of \cc{GetTimeRange()}: either the
subsequent check will succeed or it will fail.  In the case where the
subsequent check will succeed, the developer can use $\currenttime{t}$
to then open the time-limited invariant and access the $\currentepoch{e}$
resource inside it.  (\autoref{sec:case:leases} has a more detailed
discussion of how this allows proving linearizability for reads.)
When the proof eventually considers the actual comparison in the \cc{if}
statement, only one of the \cc{if} branches will be viable in each of
the two proof cases.

\begin{Subsection}{full}{Reasoning about exactly-once operations \extendedsymbol{}}
\label{sec:design:exactlyonce}

vRSM's \cc{exactlyonce} library deduplicates requests, by running the
request the first time it is seen, and using the saved reply if the
request appears again.
(This is part of the state machine and hence the reply cache is duplicated across replicas.)
This poses a proof challenge in situations
when handling the request requires exclusive ownership of resources:
on the one hand, the proof must be ready to provide these resources
in case this is the first time the request is seen (and thus must be
executed), but on the other hand, the proof cannot have a second copy
of the resources if this is a duplicate of the original request (because
the resources are exclusive and have already been used).

For instance, the top-level spec for \kv{}'s $\cc{Put}$ is written
in terms of an exclusive $\kvptsto{k}{v}$ that denotes ownership of a
particular key $k$ and that its value is $v$.  A high level proof plan
for this is to first transfer ownership of $\kvptsto{k}{v}$ to the
primary server so that it can be certain that it is safe to \cc{Put}
on the key.  However, \sys{}'s RPC library can only send duplicable
resources, because it may retransmit the request.  Replication complicates
ownership transfer even more: even if the client's \cc{Put} somehow does
transfer $\kvptsto{k}{v}$ to a primary server, that primary may crash
and reconfiguration may set up a new primary.  Then the client retries
and---following the high-level proof plan of transferring resources to
the server in charge of handling a request---the proof would need to
somehow reclaim $\kvptsto{k}{v}$ from the old primary and then transfer
it to the new primary.

To deal with such ownership transfer challenges, \sys{} allows proofs to use an
\emph{escrow} pattern~\cite{sharma:grove-thesis}.
The idea of the escrow pattern is to \emph{indirectly} transfer ownership of
some non-duplicable resources by ``depositing'' the resource in a ``known
location'' (an escrow invariant), and then only transferring duplicable
\emph{knowledge} that the deposit has happened.  For this to work, the other
party needs to have the sole right to take things out of the escrow.

\paragraph{Example.}
At the beginning of a client's \cc{Put}, the client owns $\kvptsto{k}{v}$.  To
allow a server to access it, the client gives up ownership to establish the
invariant $\knowInv{}{\kvptsto{k}{v} \vee \tok}$, called the \emph{escrow
invariant}. Here, $\tok$ is an exclusive ghost token that is initially owned by
the replica servers. Exclusive means that $\tok \ast \tok \to \FALSEsm$.  The full
proof of the \cc{exactlyonce} library deals with multiple requests and escrows
by having a separate token for each request ID, all of which are initially owned by
the replica servers.

When the client sends the \cc{Put} operation to a primary server, it only
transfers knowledge of the escrow invariant. Because knowledge is duplicable,
the client can retry and also transfer knowledge of the invariant to future
primary servers. On the other end, when a server receives a request it also
gains knowledge of the invariant. If the request is fresh, then the server will
own $\tok$. When the request is committed, the server opens the escrow invariant
and has to deal with the two possible disjuncts in the invariant. In the left
disjunct, the server now has ownership of $\kvptsto{k}{v}$ and can close the
invariant by placing $\tok$ inside of it. In the right disjunct, the proof can
derive a contradiction because there would be two copies of the exclusive
$\tok$: one from the invariant one already owned by the server.  Finally, if the
request is found to be a duplicate when committed, then the server does not own
$\tok$ to start with, and cannot get ownership of $\kvptsto{k}{v}$; instead, it
replies to the client with the previously cached reply.

\end{Subsection}

\begin{Subsection}{full}{Linearizing read operations \extendedsymbol{}}
\label{sec:design:prophecy}

When a client retransmits a read operation to vRSM, the server re-executes
it instead of using the \cc{exactlyonce} library to look up the previous
response.  This improves performance because it avoids the cost of logging
the read operation to disk.  Re-executing reads is safe in terms of the
server state: since the read has no side effects, it is safe to run any
number of reads.  Re-executed reads might return different results each
time they are executed; however, when the clerk eventually receives the
response to one of its retransmitted requests, it will use that response
as if that was the only read that was ever executed, ignoring all others.
This makes this optimization safe from the client perspective as well.

Proving correctness for this read optimization in \sys{} is challenging.
vRSM specifies linearizability for reads by establishing an exact order in which
operations are executed, requiring the proof to ``linearize'' an operation by
adding it to this execution order at most once and at the instant that the
operation logically executes.  (Specifically, vRSM is specified in the style of
logical atomicity~\cite{jacobs:logatom}.)  Importantly, proofs must add reads to
the execution order before later writes are acknowledged to clients.

The proof challenge lies in determining when to linearize a read.  On one hand,
at the moment that the read executes, the proof does not know if that
execution's response will be received by the client, so it is unclear whether to
linearize the read. On the other hand, if the proof waits until the client
receives a response, it is too late to linearize the read because concurrent
write operations might have been acknowledged to other clients and already added
to the execution order.

\sys{} enables the vRSM proof to address this challenge by providing
support for prophecy variables~\cite{jung:prophecy, abadi:prophecy}, which
was initially added to Perennial to support vMVCC~\cite{chang:vmvcc}.
Specifically, when a clerk issues a read request RPC to a vRSM server,
it uses a prophecy variable to speculate on what the eventual response
will be.  When the server executes a read, it checks whether the
read result matches the prophecy variable speculation, and if so,
whether this is the first execution of the read (using ghost state to
track re-execution).  If so, the server linearizes the read.  When the
client ultimately receives a response, it resolves the prophecy variable
prediction against the actual response.  In case the speculation
was wrong, the proof stops with a contradiction.  In case the
speculation was right, the proof learns that the read was linearized
with the expected value.

\end{Subsection}

\begin{Subsection}{full}{Reasoning about crashes \extendedsymbol{}}
\label{sec:design:chl}

\sys{} reasons about crashes by extending Crash Hoare
logic~\cite{chajed:go-journal,chen:fscq} to
the distributed setting, and borrows the notion of a
\emph{crash obligation} to encode what can be assumed
about the durable state of the system after a crash.
To reason about the state of durable storage, \sys{} introduces a file
points-to resource, written $\fileptsto{filename}{\textit{data}}{j}$, which says
that \cc{filename} on node $j$ contains \emph{data}. File points-to
resources are \emph{durable}, so ownership of them usually appears
in the crash obligation of a node.  In contrast, heap resources are
\emph{volatile} and cannot be made part of a crash obligation, because
the heap will be lost after a crash.

\paragraph{Example.}
The proof of \kv{}'s replica servers maintains the per-node crash obligation
$$\exists e, \exists \ell, \fileptsto{log.dat}{\textlogsm{toBytes}(\ell)}{j} \ast \aol{\cc{accepted}_j[e]}{\ell}.$$
The crash obligation
says that at all times, replica server $j$ must own the log file and that the
contents of the file match the list operations in the \cc{accepted} ghost
state.

The \cc{accepted} ghost state tracks all promises made to other nodes about
operations that have been applied.  The per-node crash obligation therefore
captures the need to prove that operations are made durable before being
acknowledged to other nodes.

\end{Subsection}

%% file: case.tex
\section{How \sys{} rules out bugs}
\label{sec:case}

This section sketches the specification and proof for several
components from~\autoref{sec:system}. It also poses some tricky scenarios for
the components and explains either (1) why the scenario would result in buggy
behavior and where the proof would get stuck because of the bug, or (2) why the
scenario is subtly safe and how the proof covers it.  A common theme is that the
proofs center around choosing the right kinds of resources and do not need to
break into cases for the different scenarios.

\subsection{Primary replica server}
\label{sec:case:pb}

\paragraph{Specification.}
An important part of vRSM's primary/backup replication protocol is
embodied in the primary server's \cc{Apply} function, whose job is to add
a new request to the log on the primary as well as all backup replicas.
\autoref{fig:apply} shows the simplified code for this function, and
in the rest of this subsection, we will walk through the proof of its
correctness.

The spec we aim to prove for \cc{Apply} is that its postcondition is
$\exists \ell,\, \aolb{\cc{committed}}{\ell + [\cc{op}]}$.  This is a
formal way of stating that, after \cc{Apply(op)} is done, the client's
\cc{op} definitely shows up in the committed log somewhere.  The \cc{op}
might appear in the log after a number of other operations, which may
have come from other clients.  Similarly, \cc{op} might not be the
latest operation in the log either, if other operations arrived after
it; however, the use of $\sqsupseteq$ allows the postcondition to ignore
subsequent parts of the log.
The spec does allow the operation to be added multiple times;
a stronger exactly-once spec can be obtained on top of this spec
via the \cc{exactlyonce} library.

\paragraph{Proof.}

The first line acquires the primary server lock, which serializes
concurrent calls to \cc{Apply}.  In the proof, the postcondition of
\cc{s.mutex.Lock()} provides ownership of the primary's points-to
resource, $\aol{\cc{accepted}_0[e]}{\ell}$.  While holding the lock,
the primary establishes the order in which this \cc{op} will
execute (namely, \cc{nextIndex}), and applies \cc{op} to its local
state.  At this point, the proof updates the thread's ownership of
$\aol{\cc{accepted}_0[e]}{\ell}$ to $\aol{\cc{accepted}_0[e]}{\ell + [\cc{op}]}$,
and gets the knowledge resource $\aolb{\cc{accepted}_0[e]}{\ell + [\cc{op}]}$
before releasing the lock.  To release the lock, the proof must give up
ownership of $\aol{\cc{accepted}_0[e]}{\ell + [\cc{op}]}$, but retains
knowledge of $\aolb{\cc{accepted}_0[e]}{\ell + [\cc{op}]}$, which will
be useful later on.

Next, the primary invokes \cc{ApplyAsBackupRPC} concurrently on all
of the backup servers, passing in \cc{nextIndex} to ensure that
\cc{op} is added to the backup's log at the right position.
Using the RPC spec shown at the end of \autoref{sec:design:rpc}, the primary
gets a promise that the $j$th backup accepted the operation, in the form of the lower
bound resource ${\aolb{\cc{accepted}_{j+1}[e]}{\ell + [\cc{op}]}}$. The RPC spec is
established in a separate proof.

The proof of $\cc{Apply}$ collects these postconditions of the $n$ calls
to \cc{ApplyAsBackupRPC} to get the resources
$({\aolb{\cc{accepted}_1[e]}{\ell + [\cc{op}]}})
\ast \cdots \ast
({\aolb{\cc{accepted}_n[e]}{\ell + [\cc{op}]}})$.
At this point, the proof has enough lower bound resources to be
certain that the operation is committed.  The proof opens
the invariant $I_{\textit{rep}}$ defined in \autoref{sec:design:invariants} and temporarily gets
ownership of $\aol{\cc{committed}}{\ell}$.\footnote{Strictly speaking, the
invariant refers to some $l'$, but the proof can handle the case $l'\ne l$ through
a combination of available facts and invariants.}  The proof then updates it to
$\aol{\cc{committed}}{\ell + [\cc{op}]}$.  With this points-to, the proof gets
the lower-bound resource $\aolb{\cc{committed}}{\ell + [\cc{op}]}$, which is
needed for the postcondition of \cc{Apply}.

Before the proof can complete, it must close the invariant
$I_{\textit{rep}}$ by returning ownership of the committed points-to.
To close the invariant $I_{\textit{rep}}$, the proof gives up the resources
\begin{small}
\begin{align*}
  (\aol {\cc{committed}} {\ell + [\cc{op}]}) &~\ast \\
  (\aolb{\cc{accepted}_0[e]} {\ell + [\cc{op}]}) &\ast
  \cdots \ast
  (\aolb{\cc{accepted}_n[e]} {\ell + [\cc{op}]}),
\end{align*}
\end{small}
which matches $I_{\textit{rep}}$'s body with $\ell + [\cc{op}]$ in place
of $\ell$.

Since the lower-bound resource for the committed list matches the desired
postcondition, the proof is complete.

\paragraph{What if a backup concurrently applies more operations?}
If backup $j$ applies an operation concurrently, it will end up growing
its $\cc{accepted}_j[e]$ list (that is what happens, for instance, during
\cc{ApplyAsBackupRPC}). But, since the points-to is append-only, the
lower bound resource $\aolb{\cc{accepted}_j[e]}{\ell + [\cc{op}]}$
that the primary received from calling \cc{ApplyAsBackupRPC}
is still valid, which captures the fact that the
operation \cc{op} appears in backup $j$'s log, even if it's not the
latest. Since the operation is still in all the backups' logs, it is
safe for the primary to reply \cc{OK}.

\begin{Paragraph}{full}{Why can't a backup concurrently remove operations?  \extendedsymbol{}}
From the point of view of the primary server, a (buggy) backup $j$ might
remove the latest operation \cc{op}, perhaps because it
crashed and restarted with a truncated log recovered from its disk.
In this case, it would be incorrect for the primary to reply \cc{OK}
to the client.

Fortunately, this possibility is ruled out in the proof by the resources
that the primary owns. By the time the proof of \cc{Apply} has
$\aolb{\cc{accepted}_j[e]}{\ell + [\cc{op}]}$, the proof knows that backup $j$
owned $\aol{\cc{accepted}_j[e]}{\ell + [\cc{op}]}$ at some point, and since
the points-to is append-only, it would be impossible for the list to
shrink. So, the primary does not need to consider this possibility,
and can be sure that it is safe to reply \cc{OK}.
This reasoning even works across epochs, since one of these append-only lists
will be used as the starting point for the next epoch.

Note that if the code for \cc{ApplyAsBackup} really was buggy, and could
lose operations on the backup, the developer would not be able to prove
the correctness of \cc{ApplyAsBackup}.
The append-only nature of $\cc{accepted}_j[e]$ and the backup's crash
obligation mean that
the backup's proof must ensure the log is properly
preserved under crashes and recovery.

\end{Paragraph}

\subsection{Reconfiguration}
\label{sec:case:reconfig}

A separate challenging aspect of vRSM lies in reconfiguration.
This subsection will walk through the proof of \cc{Reconfigure}, as
shown in \autoref{fig:reconfig_code}.  This function is invoked on an
operator's machine when the operator wants to change the set of servers,
perhaps adding new machines to replace failed ones.  The spec
is the following:

$$\hoareV{\text{all \cc{newServers} are valid replica servers}}
         {\cc{Reconfigure(newServers)}}
         {\top}$$

This specification states that calling \cc{Reconfigure()} requires all of
the servers specified in \cc{newServers} to be already running the vRSM
replica software (although the servers might have been just installed,
containing no key-value state), so that the reconfiguration logic knows
it is safe to use them as a replica.  \cc{Reconfigure()} will contact
both the old servers and the new servers, transferring the state to the
new servers before registering them with the configuration service.

The postcondition in \cc{Reconfigure()}'s specification appears to be
weak, in the sense that it does not promise that \cc{Reconfigure()}
will make progress.  This is because \sys{} is limited to safety
properties---ensuring that vRSM never returns the wrong result---as
opposed to liveness, such as guaranteeing that a client will receive
a response.  However, the $\top$ postcondition \emph{does} actually
guarantee an absence of all safety bugs during reconfiguration, such as
corrupting the state sent to the new servers, losing some operations
applied concurrently by the old servers, etc.  This is because
\cc{Reconfigure()} does not own any interesting resources to start with,
which precludes it from tampering with any resources held by the rest
of vRSM.  Any resources that \cc{Reconfigure()} obtains must come from
invariants, such as $I_{\textit{rep}}$.  However, \sys{} requires that
the proof correctly re-establishes any invariant that is opened, thereby
ensuring that the invariants are maintained throughout the execution
of \cc{Reconfigure()}.

\paragraph{Proof.}

The overall structure of \cc{Reconfigure()} is to get a
new epoch, then choose one of the old servers to seal and get a copy
of the old state from, then send this state to all of the new servers,
register the configuration, and activate the new primary.  The proof
relies on the fact that \sys{}'s append-only list resource, used to
represent vRSM's log, is indexed by epoch.

A key aspect to the proof lies in the resource returned to
\cc{Reconfigure()} by the \cc{GetStateAndSeal} RPC.
The postcondition of \cc{GetStateAndSeal(newEpoch)} is:
\begin{align*}
\exists \ell,\, & \left(\text{\cc{oldState} corresponds to log of operations $\ell$}\right)
\ast \\
& \, \aolro{\cc{accepted}_j[e]}{\ell}.
\end{align*}

Once the proof receives ownership of the read-only resource
$\aolro{\cc{accepted}_j[e]}{\ell}$ for the old epoch, it can conclude that
the old configuration will not apply any more operations.
This is because replica $j$ must have given up ownership of its
$\aol{\cc{accepted}_j[e]}{\ell}$ to produce the read-only resource,
which precludes it from extending $\ell$ with more operations.  After
reconfiguration completes, a new append-only list resource
$\cc{accepted}_j[e+1]$ will be allocated (assuming the new epoch is $e+1$),
which allows vRSM to append new operations.

As in the primary/backup replication proof, \sys{} allows the developer
to prove the correctness of \cc{Reconfigure()} with modular reasoning, without
having to consider explicit interleavings with other parts of the system.
Nonetheless, the proof does rule out bugs due to subtle interactions, as
follows:

\paragraph{Why can't the old primary commit additional operations?}
The developer might worry that after \cc{Reconfigure()} fetches the old
state, the old primary could execute additional operations.  This would
mean that \cc{Reconfigure()} will send an incomplete state to the new
servers, losing an operation.  This possibility is ruled out in vRSM's
proof due to the read-only resource sent back by the \cc{GetStateAndSeal}
RPC.  Having this read-only resource implies that the old primary cannot
add any more operations to $\aol{\cc{committed}}{\ell}$, since that would
require adding the operation to every old replica, which would in turn
contradict the read-only resource.

\paragraph{What if the log contains operations that were never
committed?}  It is possible that the log obtained by \cc{Reconfigure()}
contains some operations that were never committed in the old
configuration, for example if the primary sent the operation to some
but not all backups before failing.  However, it is nonetheless safe to
commit those operations in the new config.  This is because the primary
first adds an operation to its own log before sending it to the backups.
vRSM captures this in an invariant by stating that every operation in
a backup's $\cc{accepted}_j[e]$ must also appear at the same position in
the primary's $\cc{accepted}_0[e]$, making it safe to commit.  A client
clerk will learn the outcome of its operation when it resends its request
to the servers in the latest configuration.

\begin{Paragraph}{full}{What if a replica is sealed again?  \extendedsymbol{}}
If some machine initiates
reconfiguration, it might seal one of the existing replicas but then
crash.  A second reconfiguration may then try to seal the replica for
a second time.  In vRSM, a sealed replica can be sealed any number
of times in the same epoch, until a new epoch starts.  In the proof,
this is reflected by the fact that $\aolro{\cc{accepted}_j[e]}{\ell}$
is duplicable and can be sent in response to repeated seal requests.

\end{Paragraph}

\subsection{Lease-based reads}
\label{sec:case:leases}

The key challenge for lease-based reads lies in ensuring that the result
returned by \cc{ApplyReadonly()}, as shown in \autoref{fig:lease_code}, reflects
a properly linearized execution: the result cannot be
stale (i.e., missing writes that have already finished), and cannot be rolled
back (i.e., reflect uncommitted writes that might be lost due to a crash or
reconfiguration).

vRSM proves that the value is not stale by using the lease.
The read operation will be executed against the state of the local
server, represented by $\cc{accepted}_j[e]$.  At the instant of the
\cc{GetTimeRange} invocation on line 4, the proof opens the lease
invariant to obtain $\currentepoch{e}$, and opens the replication invariant
$I_{\textit{rep}}$ to obtain the fact that all of the operations in
$\cc{committed}$ are also in $\cc{accepted}_j[e]$ (which has not been
superseded by any higher epoch $e'$).

To prove the second part (that the returned value cannot observe
writes that will be rolled back), vRSM waits until all of the writes
preceding the read's linearization point to the same key have been committed,
in \cc{waitForCommitted}.  There are two cases to consider.  First, there may
be no pending writes, e.g., if at the instant of the \cc{GetTimeRange}
invocation, \cc{idx} is already committed.  In this case, the proof
linearizes the read operation immediately at the instant of \cc{GetTimeRange}.
The second possibility is that there are some pending writes to commit.
In this case, the proof maintains a set of ghost state updates (based on
the helping pattern~\cite{censor-hillel:helping,chajed:perennial})
that must be logically applied when
the preceding write is committed (which happens in the primary server's
proof of \cc{Apply()}).  Note that this case distinction is only possible
in the proof; the code does not know which case it's in
when running \cc{GetTimeRange}, so \cc{waitForCommitted} waits
for \cc{idx} to be committed in both cases (and, in the first case,
returns right away).  If reconfiguration happens in the meantime, and
the epoch number changed before \cc{idx} could be committed, the read
result might not be valid, and the server tells the client to retry.

\paragraph{What if the server pauses for a long time after checking the
lease?}
A typical concern with leases is the freshness of the lease
check.  What if the server running \cc{ApplyReadonly()} does the lease
check on line 4 of \autoref{fig:lease_code}, but then pauses for a long
time (e.g., due to garbage collection) before actually executing the read
operation on line 6?  By that time, reconfiguration may have taken place,
choosing a new primary, and that primary has executed more writes.

Such delays cannot violate correctness.  Even if a new primary executes writes,
the read
will be linearized \emph{before} those writes.
This is allowed because the read request arrived before reconfiguration
(since the lease check passed after the request was sent).  The lock
held by \cc{ApplyReadonly()} ensures that the server's state does not
change between the lease check and the execution of the read operation.

In the proof, the read operation is linearized at the instant of the
\cc{GetTimeRange()} invocation in \autoref{fig:lease_code}, if the
returned latest time is less than the lease expiration time.  This allows
the proof to open the lease invariant to establish that the
epoch is still $e$, and thus the in-memory state of that replica $j$
that will be accessed by \cc{LocalRead}, corresponding to
$\cc{accepted}_j[e]$, will be committed if \cc{waitForCommitted()}
succeeds.

\subsection{Client cache consistency with leases}
\label{sec:case:cache}

The top-level spec for the key-value caching library \cc{cachekv} is that \cc{Get}
and \cc{Put} behave like a linearizable key-value service, with
\cc{GetAndCache} working functionally like a \cc{Get}.

To prove the linearizability of its lease-based caching, the
library uses a ghost map resource, which has a $\kvptsto{k}{v}$
assertion representing the fact that the current value of $k$ is $v$.
The library maintains two invariants: one global invariant across all
instances of the key-value caching library that share the same state,
and one local lock invariant for each node's own cache, protected by a
mutex lock on that node.  The global invariant maintains, for each key,
a time-bounded invariant $\lease{L}{\kvptsto{k}{v}}$ and ownership of
the expiration time $\leaseexpiration{L}{\textit{exp}}$, corresponding
to the expiration time encoded in the underlying key-value pair.
The expiration time may be in the past if the most recent lease on
$k$ has already expired.  On each node, the library's lock invariant
maintains $\lease{L}{\kvptsto{k}{v}}$ and a lower bound on the lease
expiration time, $\leaseexpirationlb{L}{\textit{exp}}$, corresponding
to the expiration time stored in its local cache.

When a client executes \cc{Get}, the library acquires its per-node lock
and checks the lease expiration time for the requested key.  If the key is
cached and the lease is not expired, the library opens the time-bounded
invariant to prove that the cached value is the current value for that
key.  \cc{Put} waits until the lease expires, at which point its proof
can reclaim the $\kvptsto{k}{v}$ resource from the lease, update it to
the new value $v'$, and then put it back into a new lease with the same
(expired) expiration time, to re-establish the global invariant that every
key corresponds to some lease.  The proof of \cc{GetAndCache} extends
the lease in the global invariant, and makes a copy of the lease and
the corresponding expiration time lower bound in its local node invariant.

\paragraph{Why can't a \cc{Put} change a currently cached value?}
The proof of \cc{Put} has to update $\kvptsto{k}{v}$ to maintain the
invariant.  To do this update, the proof needs ownership of the
points-to.  If a client currently has that key cached (and not expired),
then the time-bounded invariant containing that key's points-to has not
expired yet, and the \cc{Put} proof cannot reclaim it.

\paragraph{Why can't \cc{GetAndCache} decrease the lease expiration
time instead of extending it?}
This would result in a bug because a \cc{Put} might change the value while
another client still believes it has an up-to-date cache.  The proof would
get stuck when re-establishing the global invariant with the decreased
lease expiration time, because that would require updating ownership of
the expiration time $\leaseexpiration{L}{\textit{exp}}$ to a smaller number
$\textit{exp}' < \textit{exp}$, which is not possible: the lease extend rule
from~\autoref{sec:design:leases} only allows the expiration time to
increase.

\begin{Subsection}{full}{Versioning read-only operations \extendedsymbol{}}
\label{sec:case:ro}

When the developer supplies a \cc{VersionedStateMachine} struct to the
vRSM library, the developer must prove that their methods satisfy the
spec expected for this interface.  To do this, the developer can choose
what resource they want to use to represent ownership of the in-memory
state on a given node, which we will denote by $\ownvsm{\textit{ops}}$.
$\textit{ops}$ refers to the list of operations that have been
executed so far, and $\ownvsm{\textit{ops}}$ says that the local state
corresponds to exactly those operations applied in that order, even if
the implementation does not keep all of them in memory.  The developer
also specifies a logical function, $\textlogsm{ComputeReply}(\textit{ops}, \textit{newop})$,
which determines the expected reply for a new operation $\textit{newop}$
executed after a list of preceding operations $\textit{ops}$.

A key optimization for achieving fast reads is to avoid waiting
for all of the writes to be committed before responding to a read.
The \cc{Read()} method in \autoref{fig:vsm_code} specifies the
write operation that the read depends on, by returning the \cc{uint64}
index of this write dependency as the first return value.  The specification
for \cc{Read()}, which the developer must prove, requires that this
index be correct:
\begin{small}
$$\hoareV{\ownvsm{\textit{ops}}}
         {\cc{Read(op)}}
         {\Ret{(\cc{idx}, \cc{res})}
          \ownvsm{\textit{ops}} \ast \textlogsm{StableReply}(\textit{ops}, \cc{idx}, \cc{op}, \cc{res})
          }$$
\end{small}
where $\textlogsm{StableReply}(\textit{ops}, \cc{idx}, \cc{op}, \cc{res})$ is defined
as:
\begin{small}
\begin{align*}
\forall \textit{ops'},\, &
          \textit{ops'} \sqsubseteq \textit{ops} \land
          \textlogsm{len}(\textit{ops'}) \geq \cc{idx} \rightarrow \\
& \, \textlogsm{ComputeReply}(\textit{ops'}, \cc{op}) = \cc{res}
\end{align*}
\end{small}
This specification says that \cc{Read()} does not modify the
logical state (the postcondition returns the same $\ownvsm{\textit{ops}}$
that it got in the precondition) and captures the necessary conditions for
the vRSM library to safely execute the read after \cc{idx} is committed.

\paragraph{Why can't the application incorrectly track read dependencies?}
If \cc{Read()} fails to correctly track read-write dependencies,
it might return an \cc{idx} value that's too small---that is,
missing a relevant recent write that affects the read result.
The \cc{Read} specification ensures this cannot happen.
$\textlogsm{StableReply}$ requires that the read \cc{op} returns the same
result \cc{res} regardless of where in the log of operations it is
executed, from \cc{idx} up to the entire current history $\textit{ops}$.
If \cc{Read()} had a bug, the developer would not be able to
prove this specification for their implementation.

\paragraph{Why can't a read-only operation have side effects?}
The implementation of \cc{Read()} is arbitrary code, and can
modify state.  It is safe for \cc{Read()} to make internal
changes, which are not visible at the interface (e.g., modifying some
internal caches).  However, if \cc{Read()} makes changes to
the state that are visible at the interface level (e.g., inadvertently
changing the value of some key in \kv{}), it would violate vRSM's
guarantees, such as linearizability and replication.  By specifying
$\ownvsm{\textit{ops}}$, the developer implicitly decides which in-memory
state is visible: state not reflected in $\textit{ops}$ is not visible to $\textlogsm{ComputeReply}$ and hence
cannot affect vRSM's behavior.  Since the postcondition of \cc{Read()}
requires $\ownvsm{\textit{ops}}$, the developer would be unable to prove
this postcondition if \cc{Read()} made any visible changes.

\end{Subsection}

\begin{Subsection}{full}{Application state consistency in \cc{paxos} \extendedsymbol{}}
\label{sec:case:configpaxos}

The proof of the \cc{paxos} library, which is used to replicate \cc{configservice},
is structured much like the proof of the primary-backup library.
One difference is that \cc{paxos} uses majority quorums instead of
replicating to every server, which means that a majority quorum $q$
of the $\cc{accepted}_i$ resources agree with the $\cc{committed}$
resource, rather than all of them:
$$
\boxed{
\begin{array}{@{}r@{~}l@{}}
\exists \ell, \exists e, \exists q, &
  \aol {\cc{committed}} {\ell} \ast
  \ulcorner 2 * \textlogsm{size}(q) > n \urcorner \ast
  \\
& \Sep_{i \in q}\left(\aolb{\cc{accepted}_i[e]} {\ell}\right)
\end{array}
}
$$

Intersection of quorums allows the proof to conclude that the new leader
has all of the previously committed operations after checking with a
majority of the $n$ servers, allowing it to gain ownership of the leader
resource for the new epoch.

The big difference from primary-backup replication is that
\cc{paxos} provides weak guarantees when reading the state.
\autoref{fig:paxosbeginspec} shows the precise specification for
\cc{Begin}, which returns two values: the \cc{oldstate} value and the
\cc{commit} callback function.  The $\paxoswf$ predicate (chosen by the caller of
the \cc{paxos} library, such as \cc{configservice}) captures the notion of
a well-formed state.  For example, in the case of \cc{configservice}, $\paxoswf$
defines a well-formed encoding of the \cc{configservice} state consisting
of an epoch, a configuration, lease expiration, etc.  \cc{paxos} captures
the fact that \cc{oldstate} might be stale by requiring that $\paxoswf$ is
duplicable: that is, $\paxoswf$ can only capture properties that are always going
to be true about a state, and cannot capture non-duplicable facts such
as freshness.

\begin{figure}[ht]
\vspace{-\baselineskip}
$$
  \hoareV[c]
    { \top }
    { \cc{paxos.Begin()} }
    { \begin{aligned}[c]
      \Ret{(\cc{oldstate}, \cc{commit})}
      & \paxoswf(\cc{oldstate}) \ast \\
      & \textlogsm{commitspec}(\cc{commit}, \cc{oldstate})
      \end{aligned}
    }
$$
    \caption{Specification for \cc{Begin()} in the \cc{paxos} library.
      The definition of $\textlogsm{commitspec}$ is shown in
      \autoref{fig:paxoscommitspec}.
    }
    \label{fig:paxosbeginspec}
\end{figure}

The second component of the postcondition, $\textlogsm{commitspec}$,
allows the caller to later attempt to commit a new state by
invoking the callback \cc{commit}, passing the new state as an
argument.  The definition of $\textlogsm{commitspec}$ is shown
in \autoref{fig:paxoscommitspec}.  Its precondition requires that
\cc{newstate} is well-formed according to $\paxoswf$.

\begin{figure}[ht]
$$
\begin{aligned}[c]
& \forall \cc{newstate}, \\
& \hoareV[c]
    { \paxoswf(\cc{newstate}) }
    { \ahoareV[c]
      { \forall\forall \textit{state'}, \paxosstate{\textit{state'}} }
      { \cc{commit}(\cc{newstate}) }
      { \begin{aligned}[c]
          \exists \cc{err}, \text{if } \cc{err} & \text{ then } \paxosstate{\textit{state'}} \\
          & \text{ else }
            \begin{aligned}[t]
            & \ulcorner \textit{state'} = \cc{oldstate} \urcorner ~\ast \\
            & \paxosstate{\cc{newstate}}
            \end{aligned}
          \end{aligned}
      }
    }
    { \Ret{\cc{err}} \top }
\end{aligned}
$$
  \caption{Definition of $\textlogsm{commitspec}(\cc{commit}, \cc{oldstate})$.}
  \label{fig:paxoscommitspec}
\end{figure}

The rest of the $\textlogsm{commitspec}$ specification is written using
angle brackets, which in Grove (like in Iris) indicates a \emph{logically
atomic specification}~\cite{jung:prophecy}.  This means that the
transition between the angle-bracketed pre- and postcondition occurs
atomically at the \emph{linearization point} inside this function,
which---crucially---lets multiple threads or nodes call this operation
concurrently on the same exclusive resources.

In this case, that exclusive resource is $\paxosstate{\textit{state'}}$, which
represents ownership of the fact that $\textit{state'}$ is the latest
committed state.  This resource is held by the application's invariant;
for instance, \cc{configservice} maintains an invariant owning this
resource and specifying that there is also a lease valid until the
expiration time encoded in that state.

The \cc{commit} specification now says that, at some point during \cc{commit}'s
execution (specifically, at the linearization point),
it will need ownership of the $\paxosstate{\textit{state'}}$
resource, for some $\textit{state'}$.
(The special $\forall\forall$ quantifier indicates that this variable is quantified across both the pre- and postcondition,
and that its value only gets determined at the linearization point. The value is chosen by the client that invokes \cc{commit}.)
This $\textit{state'}$ might not
be the same as \cc{oldstate}, which captures the possibility that other
transactions may have modified the state by the time \cc{commit}
is executed.  The logical atomicity postcondition says that, either
\cc{commit} will return an error and the state remains $\textit{state'}$, or
\cc{commit} will return success, in which case the caller learns that $\textit{state'}$
was in fact the same as \cc{oldstate} and the committed state is changed to
\cc{newstate}.

\paragraph{What if \cc{configservice} gets stale or invalid state from
\cc{paxos}?}
When \cc{configservice}'s implementation of \cc{ReserveEpochAndGetConfig},
shown in \autoref{fig:configimpl_code}, gets the starting state
\cc{oldstate}, \cc{Begin}'s postcondition promises that it is well-formed
according to $\paxoswf$, which allows the proof to conclude that it's
safe to unmarshal it.  However, because $\paxoswf$ is duplicable,
the proof would not be able to (incorrectly) conclude anything about
the freshness of \cc{oldstate}, such as whether the corresponding
lease is currently valid.  When invoking \cc{commit}, if \cc{commit}
is about to return success, then the logically
atomic specification allows the proof to conclude that the state was
indeed fresh.  However,
the precondition of \cc{commit} requires the caller (the proof of
\cc{ReserveEpochAndGetConfig}) to establish $\paxoswf(\cc{newstate})$
before knowing whether \cc{oldstate} was fresh or not, which
prevents the proof from embedding any information about
whether \cc{oldstate} was ever committed into $\paxoswf(\cc{newstate})$.

\end{Subsection}

\begin{Subsection}{full}{Bank transactions \extendedsymbol{}}
\label{sec:case:bank}

The bank uses one instance of \kv{} to store account balances and one
instance of a lock service to maintain locks on individual accounts.
The specification for the lock service allows maintaining a lock invariant
for each lock.  For each account maintained by the bank, the bank puts
the account's balance resource, $\kvptsto{\textit{acct}}{\textit{bal}}$,
in the lock invariant for lock $\textit{acct}$ maintained by the lock
service.  When the bank needs to access a specific account (either for
\cc{Transfer} or \cc{Audit}), it acquires the account's lock, which gives it
that account's points-to resource, allowing it to access the balance in \kv{}.

\paragraph{Why can't the lock service lose its state after a crash?}  If
the lock service lost track of held locks, and allowed acquiring a lock
that was already held before the lock service crashed, the proof of the
lock service would get stuck.  Specifically, the lock service would need
to give the caller access to the lock's invariant (the balance points-to
in the case of the bank example), but the lock service already handed
out that resource before the crash, and separation logic does not allow
duplicating resources.  Thus, the proof must ensure that at most one copy
of the lock resource is handed out at any given time.

\paragraph{Why can't the bank fail to follow the locking rules?}  If
the bank fails to acquire the appropriate locks, it might have race
conditions updating the same accounts from different threads.  However,
separation logic rules prevent this: accessing a key-value pair in \kv{}
requires ownership of the corresponding points-to resource for that key.
The bank has exactly one points-to resource for every key, and that
resource is stored in the lock service.  Thus, if the bank does not
acquire the lock for an account, it will not have the resource needed
to allow its proof to access the balance.

\end{Subsection}

%% file: modularity.tex
\section{Developing \sys{} proofs}
\label{sec:modularity}

A major benefit of \sys{}'s use of concurrent separation logic is that it
allows proving each function---and even each line of code---in isolation,
without explicitly considering the interleavings due to concurrency,
crashes, recovery, etc.  This not only allows incrementally proving
a single library, but also enables combining the proofs of multiple
components or functions into a single proof for the composed system.

Composability of proofs is a powerful property that is not true in
the general case.  For example, if a system is running both a key-value
service and a lock service, the key-value library might inadvertently send
a message to the lock service that causes it to release a lock, thereby
invalidating its proofs.  As another example, if the developer adds a
new RPC method to a key-value server, and this RPC incorrectly updates
data structures used by other RPCs, adding this RPC would invalidate the
proofs of existing RPC methods.  Concurrent separation logic enforces
ownership rules to ensure that all verified code is ``well-behaved'' in a
way that avoids problems like the above, and allows for sound composition.

The rest of this section presents several case studies that illustrate
the benefits of concurrent separation logic in \sys{}.

\subsection{Proving top-level spec for \kv{}}
\label{sec:modularity:top}

The top-level theorems for \kv{} are specifications for the top-level functions:

\begin{CompactEnumerate}
  \item The replica server \cc{main()} function is
    crash-idempotent~\cite{chajed:perennial, sharma:grove-thesis}.
    This covers the execution of any code invoked by
    \cc{main()}, including any RPC handlers that \cc{main()} sets up.

  \item \cc{Reconfigure(newServers)} is always safe to run, if
    \cc{newServers} are all valid replica servers.

  \item \cc{MakeClerk(configAddrs)} correctly initializes a clerk, if
    \cc{configAddrs} are the addresses of the \cc{configservice}.

  \item A \cc{clerk}'s \cc{Put(k,v)}, \cc{Get(k)}, and \cc{CondPut(k,e,v)}
    functions behave as though accessing a local in-memory key-value map with linearizable
    operations.
\end{CompactEnumerate}

Proving these theorems involves proving specs for functions in \kv{} one-by-one,
using specs for lower-level components to verify higher-level code, culminating
in a proof of the top-level functions. A client application that uses \kv{}
(e.g. the \cc{bank}) can then be verified by applying these theorems to reason
about \cc{Put} and \cc{Get} calls.

\subsection{Evolving \kv{} to add leases}

We originally built and verified the vRSM
library and \kv{} without leases. This original version
executed \cc{Get} operations as a read-write operation; that is,
by replicating the \cc{Get} operation to the primary and all backups
(including waiting for the replicas before replying to the client).

We later decided to improve performance of read-only operations by adding
leases. This involved several changes:
(1) adding \cc{GetLease} to the configuration service and making sure
other RPCs wait for any outstanding leases to expire before advancing
the epoch number;
(2) adding \cc{ApplyReadonly} to the vRSM library as well as a helper
thread on each replica that extends its lease with the \cc{GetLease} RPC;
(3) propagating the number of committed operations from the primary to
the backups;
(4) introducing \cc{VersionedStateMachine} as the vRSM library's interface
to allow some reads to happen without waiting for ongoing writes to finish;
and (5) bypassing the exactly-once operations library for
read-only operations.

In the proof before adding leases, the configuration service
always owned the epoch number and could always advance it. With leases,
ownership may reside in a time-bounded invariant, so the
proof now must establish ownership by using the fact that the code
checks for lease expiration, which allows the proof to use the time-bounded
invariant expiration rule from~\ref{sec:design:leases}. To reason about
linearizability of lease-based reads from replica servers, the proof
of the primary/backup replication and reconfiguration protocol remained
the same, but we added a new proof on top that shows that ownership of
the current epoch means any replica's state is at least as up-to-date
as the committed operations.

\begin{Subsection}{full}{Evolving \cc{configservice} to use \cc{paxos} \extendedsymbol{}}
\label{sec:modularity:configpaxos}

We originally built and verified vRSM with an in-memory and unreplicated
\cc{configservice}. This original version was not fault-tolerant, and the
previous top-level theorems of \kv{} assumed that the sole \cc{configservice}
server never crashes and restarts.  To improve the fault tolerance of the overall
vRSM system, we implemented the \cc{paxos} library and modified the
\cc{configservice} to use \cc{paxos} to manage state.

The proof of \cc{paxos} borrows heavily from the proof of primary/backup
replication and reconfiguration. We started by making a copy of the proof of the
primary/backup replication protocol, then modified key invariants to make them
quorum-based (such as the replication invariant described
in~\xref{sec:case:configpaxos}), and finally reproved many of the same lemmas
against these modified invariants.

The old \cc{configservice} used a lock to coordinate concurrent RPCs accessing
the configuration state. The new \cc{paxos}-based \cc{configservice} replaces
those \cc{Lock} and \cc{Unlock} calls with calls to \cc{paxos.Begin} and
\cc{commit} respectively, with additional error handling since \cc{commit} can
fail. Across this change, \cc{configservice}'s clerk API and specification
(\autoref{fig:config_code}) remained unchanged, except for clerks now taking
multiple network addresses as input for the multiple \cc{configservice} servers.
Since the proofs of other components that use \cc{configservice} depend only on
the specification of \cc{configservice}---and not on how that specification is
proven---those proofs also still worked without major changes.  For instance,
the new proof of \cc{replica} differs from the old proof by 18 lines added and 4
lines changed, which were largely needed because \cc{replica} servers now take
as input a list of multiple network addresses used to contact the
\cc{configservice} instead of a single address.

\end{Subsection}

\begin{Subsection}{full}{Line-by-line reasoning \extendedsymbol{}}

A developer in \sys{} can verify each component's implementation
line-by-line.  Instead of having to explicitly worry about
interleavings and interactions (like the mover reasoning in
IronFleet~\cite{hawblitzel:ironfleet}, CSpec~\cite{chajed:cspec},
and Armada~\cite{lorch:armada}), \sys{}'s resources (either owned by
specific threads or by specific invariants) indirectly constrain how
different components can interact with one another.

For example, the proofs of different functions shown in
\autoref{sec:case:pb}, \autoref{sec:case:reconfig}, and
\autoref{sec:case:leases} form a proof about the combined system,
without any explicit consideration of how these functions (\cc{Apply},
\cc{Reconfigure}, and \cc{ApplyReadonly}) will interleave.
The proof of \cc{Apply} (\autoref{sec:case:pb}) uses the $I_{\textit{rep}}$
invariant, and the proof of \cc{ApplyReadonly} also uses the same
invariant, yet the two proofs do not explicitly reason about each other
(and moreover, the proof of \cc{Apply} does not even know about the
existence of the time-bounded lease invariant).  Although the proofs are
independent, they nonetheless cover all of the possible interactions in
the resulting system.

\end{Subsection}

\begin{Subsection}{full}{Separate proofs of components \extendedsymbol{}}

The bank exemplifies how \sys{} enables verifying crash-safe,
high-performance, distributed applications out of individual verified
components.  The bank code and proof refer only to the simple interfaces
and specifications provided by the \kv{} and lock service clerks.
Despite that, the bank's specification provides a strong guarantee in
the face of crashes, reconfigurations, concurrency, network partitions,
loosely-coupled clocks, etc.  \sys{}'s modular specs and proofs also
allow concurrent development and replacement of components: the bank
application developer can write and prove their code in parallel with
the \kv{} developers, once they agree on the specification for the \kv{}
interface, and similarly, \kv{}'s implementation can be replaced with a
different key-value store (e.g., using Raft~\cite{ongaro:raft}) as long
as it satisfies the same API spec.

\end{Subsection}

%% file: impl.tex
\section{Implementation}
\label{sec:impl}

\sys{} is implemented by extending Perennial~\cite{chajed:perennial},
which is based on Iris~\cite{jung:iris-1,krebbers:ipm,jung:iris-jfp}
and Coq~\cite{coq}.  \sys{} inherits reasoning principles for concurrent
Go code from Perennial, inherits general support for interactive separation logic reasoning and ghost resources from Iris,
and adds support for distributed systems
with new reasoning principles for the network, clock, and independent
node crashes.  \sys{}'s extensions to Perennial involved \loc{1597}
lines of Coq proof for new reasoning principles, along with other
hard-to-quantify minor changes throughout Perennial.  \sys{} comes with
a distributed composition soundness theorem, which proves correctness
of \sys{}'s reasoning principles by showing that they imply a simple
statement about the behavior of the distributed system under \sys{}'s
execution model.


\begin{figure}
  \centering
  \small
  \begin{tabular}{lrrr}
    \toprule
    \bf Component &  \bf Code & \bf Spec and Proof \\
    \midrule
    \input{loc_table}
  \end{tabular}
  \vspace{-.5\baselineskip}
  \caption{Lines of Go code and Coq spec/proof for the verified components.}
  \label{fig:linecounts}
\end{figure}

\autoref{fig:linecounts} shows the breakdown of code and proof for the
different components.  The top-level specification of the bank,
which builds on most of the other components, is 52 lines.  The
specification for \kv{}, consisting of the four parts
described in \autoref{sec:modularity:top}, is 49 lines.
We confirm that the proof is complete using \cc{Print Assumptions} in Coq.
Across the different components, verification required 12$\times$ the
lines of proof as lines of code, which is comparable to other concurrent
and distributed systems verification projects: IronFleet's overhead is
slightly lower~\cite{hawblitzel:ironfleet} (and IronFleet also includes a
proof of liveness, though it does not handle thread concurrency, leases,
crashes, or reconfiguration), but GoJournal's is slightly higher~\cite{chajed:go-journal}.
One conclusion is that verifying a complete distributed system, such as
\kv{}, which handles node-local concurrency, crash recovery, leases, and
reconfiguration, did not come at the cost of an inflated proof overhead.

%% file: loc_table.tex
    {\cc{bank}}        & \loc{99}     & \loc{799} \\
    {\cc{lockservice}}        & \loc{19}     & \loc{133} \\
    {\cc{cachekv}}      & \loc{86}      & \loc{569} \\
    {\kv}      & \loc{233}      & \loc{1574} \\
    {\cc{exactlyonce}} & \loc{127}      & \loc{2272} \\
    {\cc{clerk}}                   & \loc{146}      & \loc{935} \\
    {\cc{storage}}            & \loc{227}     & \loc{3057} \\
    {\cc{configservice}}    & \loc{200}    & \loc{2797} \\
    {\cc{paxos}}    & \loc{492}    & \loc{5600} \\
    {\cc{reconfig}}         & \loc{65}  & \loc{817} \\
    {\cc{replica}}          & \loc{578}     & \loc{8093} \\
    {Time-bounded invariants} &     -- & \loc{168} \\
    {\cc{rpc}}             & \loc{163}     & \loc{1263} \\
    {Network library}         & \loc{120}     & Trusted \\
    {Filesystem library}      & \loc{50}    & Trusted \\
    \midrule
    Total & \loc{2605} & -- \\
    Total verified & \loc{2435} & \loc{28077} \\
    \bottomrule

%% file: eval.tex
\section{Evaluation}
\label{sec:eval}

To demonstrate that \sys{} is capable of verifying realistic
high-performance distributed systems, this section experimentally
demonstrates that the \kv{} prototype, which we verified using \sys{},
is able to achieve high performance.  We also demonstrate that leases are
particularly important for achieving high performance for reads in \kv{}.

\paragraph{Experimental setup.}

To evaluate \kv's performance, we use 8 CloudLab servers,
with up to 3 for replicas, 4 for clients, and 1 for
the configuration service. Each machine has an Intel Xeon CPU E5-2630v3 2.4GHz
processor with 8 cores, 64GB of RAM, an Intel 200GB 6Gb/s SSD (SSDSC2BX200G4R) for
storage, and an I350 Gigabit network card.

We generate requests using YCSB~\cite{cooper:ycsb} with uniformly random
keys and 128-byte values.  Clients run in a closed loop, issuing a new
request as soon as the previous request completes; for each data
point, we warm up the system for 20 seconds and then measure the performance for
1 minute.  To measure throughput, we keep increasing the number of clients until
the total throughput of all of the clients stops growing.

\paragraph{Baseline performance.}

To demonstrate that \kv{} achieves good performance, we compare
with Redis.  Redis is a widely-used high-performance key-value server,
written in C.  Redis targets somewhat different goals than \kv{} (it
is designed to run on a single core, it does not support synchronous
replication or live reconfiguration, etc), but it nonetheless provides a
reference point in terms of absolute performance for a key-value store.
To make Redis comparable to \kv{} in terms of its guarantees, we run
Redis with the \cc{appendfsync always} option to ensure it made changes
durable before replying, and we run \kv{} on a single core (disabling
all other cores in Linux) and with no backup replicas.
Note that Redis does not implement exactly-once semantics for its operations (if
a write gets retransmitted, it may end up being executed twice), whereas \kv{}
stores a 16-byte request ID for each operation.

\autoref{fig:redis} compares the performance of \kv{} with that of Redis.
We report the mean of 10 runs; Redis's standard deviation is 1--2\%, and
\kv{}'s is 7--11\%, due to the high variance of the Go runtime
when running on a single core.
When running on multiple cores, \kv{} achieves higher throughput---e.g.,
5.1$\times$ on 8 cores for YCSB 5\% writes, with minimal performance
variability.
The results show that \kv{}'s throughput is 67--73\% of Redis's,
and its request latency is comparable.

\begin{figure}[ht]
\centering
\small
\begin{tabular}{@{~}l@{~ ~}r@{~ ~}r@{~}}
\toprule
\bf Benchmark & \bf Redis & \bf \kv{} \\
\midrule
Throughput for YCSB 100\% writes & 99,066~req/s & 67,360~req/s \\
Throughput for YCSB 50\% writes & 107,028~req/s & 75,174~req/s \\
Throughput for YCSB 5\% writes & 118,594~req/s & 87,634~req/s \\
Read latency under low load & 81~us & 126~us \\
Write latency under low load & 538~us & 603~us \\
\bottomrule
\end{tabular}
\caption{Throughput and latency of \kv{} compared to Redis.}
\label{fig:redis}
\end{figure}

\paragraph{Reconfiguration.}

To demonstrate that \kv{} can recover from server failures by
reconfiguring the system to add new servers, all while continuing to
correctly handle client requests, we run a two-server configuration
of \kv{}.  At 10 seconds into the experiment, the primary server is
killed, and reconfiguration starts (changing to a new primary and a new
backup server).  We use a variation of the YCSB workload, with 100
clients always issuing writes, and 100 clients always issuing reads
(rather than each client issuing a mix).  This is because,
during reconfiguration, writes block if one of the servers is sealed
(which would ultimately cause all clients to block if they were
issuing reads and writes), but reads can proceed (so clients that never
issue writes can proceed).

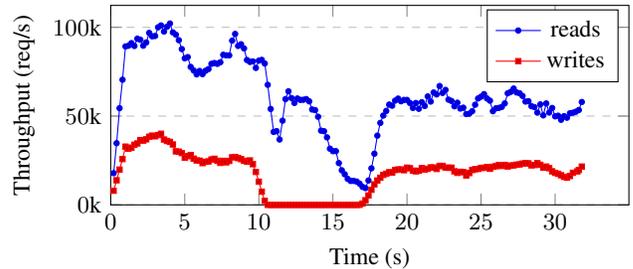
\begin{figure}[ht]
\begin{tikzpicture}
\begin{axis}[
    width=\columnwidth,
    height=0.5\columnwidth,
    xlabel={Time (s)},
    ylabel={Throughput (req/s)},
    xmin=0.0,
    ymin=0,
    scaled y ticks=base 10:-3,
    ytick scale label code/.code={},
    yticklabel={\pgfmathprintnumber{\tick}k},
    ymajorgrids=true,
    grid style=dashed,
    mark size=1pt,
]
\addplot table[col sep=comma] {data/reconfig_reads.dat};
\addplot table[col sep=comma] {data/reconfig_writes.dat};
\legend{reads,writes}
 \end{axis}
\end{tikzpicture}
\vspace{-0.7\baselineskip}
\caption{Throughput over time (averaged over 0.5 second time slices), with the
  primary crashing at 10 seconds, followed immediately by a reconfiguration to a
  new primary and backup.}%
\label{fig:eval-reconfig}
\end{figure}

\autoref{fig:eval-reconfig} shows the observed throughput by the read and write
clients over time during this experiment.  The results show that \kv{} can
continue serving reads while reconfiguring.  When the primary is initially
killed, read throughput dips while clients with outstanding read requests sent
to the primary wait to discover their connection is closed and while the remaining
backup server marshals its key-value state to be sent to the new servers.  After
the backup is done marshaling its state, and after the clients connect to the
backup and retransmit their requests, reads recover some of the throughput.
Reads do not recover to their original throughput because of stuck reads: for a
client that tries to read one of the keys whose write was in flight when the
primary was killed, \cc{waitForCommitted} returns only after
reconfiguration, because the old primary did not commit those writes before
being killed (and the backup doesn't know yet that those writes will not be
committed by the primary). As more read clients get stuck, read throughput
starts declining again.  After the state is transferred to new servers (copying
1M key-value pairs, each 128 bytes long), the system switches to the new
configuration and resumes executing reads and writes (including all
previously-stuck operations).  Most of the reconfiguration time is spent
marshalling the state and sending it to new servers via the reconfiguration
process, ($\sim$4 seconds in total).

\paragraph{Read performance with leases.}

\begin{figure}[ht]
\begin{tikzpicture}
\begin{axis}[
    width=0.99\columnwidth,
    height=0.5\columnwidth,
    xlabel={Number of servers},
    ylabel={Throughput (req/s)},
    ymin=0,
    xtick=data,
    scaled y ticks=base 10:-3,
    ytick scale label code/.code={},
    yticklabel={\pgfmathprintnumber{\tick}k},
    ytick distance=200000,
    legend style={at={(0.15, 0.9)},anchor=north west},
    grid style=dashed,
]
\addplot table[col sep=comma] {data/servers100.dat};
\addplot table[col sep=comma] {data/servers95.dat};
\addplot table[col sep=comma] {data/servers50.dat};
\addplot table[col sep=comma] {data/servers0.dat};
\pgfplotsset{domain = 1 : 8 }
\legend{0\%, 5\%, 50\%, 100\%}
\end{axis}
\end{tikzpicture}
\vspace{-1.5\baselineskip}
\caption{Peak throughput of \kv{} with increasing number of servers, labeled
  by the percentage of write operations.}%
\label{fig:scale-reads}
\end{figure}
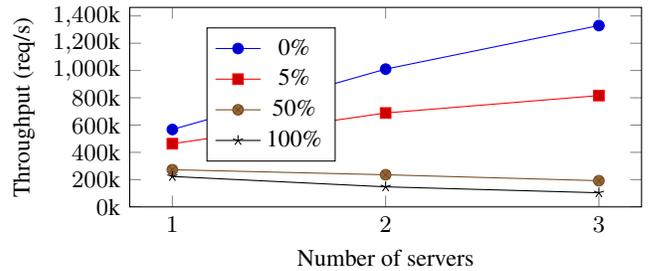

\autoref{fig:scale-reads} shows \kv{}'s throughput for different
workloads as more replicas are added.
For write-heavy workloads (50\% or 100\% writes), adding replicas reduces
performance because writes encounter more overhead at the primary server,
and there are not enough reads handled by other replicas to offset the
costs.
For read-heavy workloads, adding replicas improves performance---e.g.,
for YCSB 5\% and 0\% writes, 3 servers achieve 1.7$\times$ and 2.3$\times$
the throughput of a single server, respectively.

%% file: related.tex
\section{Related work}
\label{sec:related}

\sys{} is the first to support verifying distributed systems with
thread- and node-level concurrency, crash recovery with durable state,
time-based leases, and reconfiguration.  Verifying all of these aspects
in a single framework is critical because subtle bugs can occur due
to interactions between these features.  \kv{}, a realistic replicated
key-value store, demonstrates the benefits of \sys{}'s modular reasoning
by proving the correctness of its primary/backup replication, durable
storage, reconfiguration, concurrency, and leases.
\kv{}'s design is not novel, but rather a case study of what it takes
to build a fault-tolerant primary-backup replication system.  In doing
so, it captures key challenges in
state-of-the-art (unverified) distributed systems with primary/backup
replication and a configuration service for reconfiguration,
such as Chain Replication~\cite{vanrenesse:chain-replication},
FaRM~\cite{dragojevic:farm}, Boxwood~\cite{maccormick:boxwood},
Bigtable~\cite{chang:bigtable}, Megastore~\cite{baker:megastore},
FoundationDB~\cite{zhou:foundationdb}, Kafka~\cite{kafka}, and
Tuba~\cite{ardekani:tuba}.

\paragraph{Concurrent separation logic for distributed systems.}
Broadly similar to our work, Disel~\cite{sergey:disel} and
Aneris~\cite{krogh-jespersen:aneris, gondelman:aneris-causal}
also use concurrent separation logic in the context of distributed
systems.  However, neither Disel nor Aneris provide support for
reasoning about time-based leases or recovery from crashes, and
they have not been used to verify a system with reconfiguration.
These restrictions limit the distributed systems they can reason about.
For example, \citet{gondelman:aneris-rpc-kv} use Aneris to verify
an eventually-consistent primary-backup key-value store.  However,
that system does not support reconfiguration, so if the primary fails,
the system cannot process any further writes.  Furthermore, writes are
only lazily copied to replicas for availability, and thus reads
from replicas may return stale values.  \kv uses a combination
of reconfiguration and leases for availability when a primary fails,
while also guaranteeing that reads from replicas are up-to-date.

\paragraph{State-machine refinement.}
An alternative approach to verifying distributed systems is to prove
refinements from a high-level protocol description down to
executable code, as in IronFleet~\citep{hawblitzel:ironfleet},
Verdi~\citep{wilcox:verdi}, and IronSync~\cite{hance:ironsync}.  However,
these systems do not reason about time-based leases, reconfiguration, or node recovery.  State
machines also make it challenging to compose larger systems out of smaller
components, which features extensively in our case study.
IronSync~\cite{hance:ironsync} shows how to bring some benefits of
ownership-based reasoning to state-machine approaches,
but at a coarse granularity.

\paragraph{Distributed system abstractions.}
Adore~\citep{honore:adore} proposes an abstraction for reasoning about
reconfiguration for replicated state machine protocols, such as Raft.
\kv{}'s primary/backup replication and reconfiguration uses
a configuration service to simplify the protocol, but verifies many of the
same issues, such as concurrent request execution during reconfiguration.
\kv{} also handles interactions between reconfiguration and
crashes, recovery, leases, and thread-level concurrency, which the Adore
abstraction does not directly address.

\paragraph{Protocol reasoning.}

\TLA{}~\citep{lamport:tla, lamport:tlaplus} provides a modeling language
for concisely describing distributed protocols, which can then be
model-checked or interactively verified.  In other tools, constraining
the modeling language used for expressing protocols enables automatic or
semi-automatic proofs of correctness, such as ByMC~\citep{konnov:bymc},
Ivy~\citep{mcmillan:ivy, padon:paxos-epr}, and I4~\citep{ma:i4} and
its follow-ons.  Although protocol verification can ensure the absence
of bugs in the protocol design, many bugs in distributed systems only
manifest at the level of implementations, and so fall outside the scope
of protocol verification. \sys aims to verify implementations of systems
to address these bugs.

%% file: conclusion.tex
\section{Conclusion}
\label{sec:conclusion}

\sys{} is a library for verifying distributed systems using concurrent
separation logic (CSL).  \sys{} generalizes CSL to support distributed
systems with RPCs, leases, replication, reconfiguration, and crash
recovery.  We demonstrate \sys{} by implementing and verifying a range of
distributed system components, such as primary-backup replication,
locking, client caching, and a configuration service.  Verifying these
components in \sys{} eliminates broad classes of bugs, and comes with a
12$\times$ proof-to-code ratio, in line with previous efforts to verify
concurrent and distributed systems.  \kv{}, a key-value store built out of
these components, supports primary-backup replication and reconfiguration,
achieves 67-73\% the throughput of Redis on a single core, and scales
read throughput with more replicas due to its use of leases.

%% file: ack.tex
\subsection*{Acknowledgments}

Thanks to the anonymous reviewers, Jay Lorch, members of the MIT PDOS
group, and our shepherd, Chris Hawblitzel, for feedback that improved this
paper.  This work was supported by NSF awards CCF-2123864 and CCF-2318722.